\documentclass[a4paper, 11pt]{article}

\usepackage{amsmath, amssymb, amsfonts}  % Math stuff
\usepackage[mathscr]{eucal}
\usepackage{cases}

\usepackage{bm}
\usepackage{bbm}
\usepackage{dsfont}
\usepackage{xcolor}
\usepackage{ifpdf}

\usepackage{graphics} % for pdf, bitmapped graphics files
\usepackage{graphicx}           % For including graphics
\graphicspath{{./Images/}}

\usepackage{enumitem}
\setlist[description]{leftmargin=\parindent}

\usepackage{subfigure}

\usepackage{ifpdf}

\newcommand*\mcap{\mathbin{\mathpalette\mcapinn\relax}}
\newcommand*\mcapinn[2]{\vcenter{\hbox{$\mathsurround=0pt
  \ifx\displaystyle#1\textstyle\else#1\fi\bigcap$}}}

\newcommand*\mcup{\mathbin{\mathpalette\mcupinn\relax}}
\newcommand*\mcupinn[2]{\vcenter{\hbox{$\mathsurround=0pt
  \ifx\displaystyle#1\textstyle\else#1\fi\bigcup$}}}

\DeclareFontFamily{OT1}{pzc}{}
\DeclareFontShape{OT1}{pzc}{m}{it}{<-> s * [1.200] pzcmi7t}{}
\DeclareMathAlphabet{\mathpzc}{OT1}{pzc}{m}{it}

\usepackage{theorem}
\newtheorem{theorem}{Theorem}
\newtheorem{definition}{Definition}

\newtheorem{lemma}{Lemma}

{\theorembodyfont{\upshape}
	\newtheorem{remark}{Remark}
	\newtheorem{example}{Example}
	
}

\newcommand{\beq}{\begin{equation}}
\newcommand{\eeq}{\end{equation}}
\newcommand{\beqa}{\begin{eqnarray}}
\newcommand{\eeqa}{\end{eqnarray}}
\newcommand{\beqan}{\begin{eqnarray*}}
\newcommand{\eeqan}{\end{eqnarray*}}

\newcommand{\bite}{\begin{itemize}}
\newcommand{\eite}{\end{itemize}}
\newcommand{\benu}{\begin{enumerate}}
\newcommand{\eenu}{\end{enumerate}}

\newcommand{\R}{\mathbb{R}}

\title{\LARGE \bf Interval Consensus for Multiagent Networks\thanks{Work supported in part by a grant from the Swedish Research Council (grant n. 2015-04390 to C.A.)}}

\author{Angela Fontan\thanks{A. Fontan is  with the Division of Automatic Control, Department of Electrical Engineering, Link\"{o}ping University, SE-58183 Link\"{o}ping, Sweden.
		E-mail:  angela.fontan@liu.se}, Guodong Shi\thanks{G. Shi is with the Research School of Engineering, The Australian National University, ACT 0200,
Canberra, Australia. Email: guodong.shi@anu.edu.au}
, Xiaoming Hu\thanks{X. Hu is with the Department of Optimization and Systems Theory, Royal Institute of Technology
SE-100 44 Stockholm, Sweden. E-mail: hu@kth.se}, and Claudio Altafini\thanks{C. Altafini is  with the Division of Automatic Control, Department of Electrical Engineering, Link\"{o}ping University, SE-58183 Link\"{o}ping, Sweden.
		E-mail:   claudio.altafini@liu.se}
		}
\date{}

\begin{document}

\maketitle

\begin{abstract}
The constrained consensus problem considered in this paper, denoted interval consensus, is characterized by the fact that each agent can impose a lower and upper bound on the achievable consensus value.
Such constraints can be encoded in the consensus dynamics by saturating the values that an agent transmits to its neighboring nodes.
We show in the paper that when the intersection of the intervals imposed by the agents is nonempty, the resulting constrained consensus problem must converge to a common value inside that intersection.
In our algorithm, convergence happens in a fully distributed manner, and without need of sharing any information on the individual constraining intervals. When the intersection of the intervals is an empty set, the intrinsic nonlinearity of the network dynamics raises new challenges in understanding the node state evolution. Using Brouwer fixed-point theorem we prove that in that case there exists at least one equilibrium, and in fact the possible equilibria are locally stable if the constraints are satisfied or dissatisfied at the same time among all nodes. For graphs with sufficient sparsity it is further proven that there is a unique equilibrium that is globally attractive if the constraint intervals are pairwise disjoint.
\end{abstract}

%%%%%%%%%%%%%%%%%%%%%%%%%%%%%%%%%%%%%%%%%%

\section{Introduction}
The basic idea of a consensus problem is to achieve an agreement among a group of agents through a distributed dynamical system, encoding the values that the agents want to contribute as initial conditions of a Laplacian-like system which represents the exchanges of information among the first neighbors of a communication graph.
Owing to the Laplacian structure of the dynamics, each agent is driven only by relative states, i.e., differences between its own state and that of its neighbors.
Various algorithms have been developed using this scheme.
For instance, the average consensus problem consists of computing the average of such initial conditions, see \cite{Olfati2003Consensus}.
In a leader-follower scenario, instead, only the initial conditions of the leaders matter, and provide the values to which the followers converge, see \cite{bib:HongHuGao2006}. %\cite{Hong2006Tracking}
In a max consensus problem, the agents determine the max of their initial conditions, and all settle to that value, see \cite{bib:ShiXiaJohansson2015}.
When cooperation and competition among the agents coexist, a bipartite consensus can be achieved, provided that the graph is structurally balanced, see \cite{Altafini2013Consensus}.

In all these protocols, an agent has no authority to veto certain values of consensus, or to impose that the consensus is restricted within an admissible region.
This is a drastic limitation in certain  contexts.
For instance, in a network of processors trying to agree on sharing a computational load, each processor might have constraints on the computational resources allocable to the shared task, and accept only consensus values which are within that range.
In an opinion dynamics context, an agent might agree on a common opinion only if this is not too extreme.
In a robotic formation, a robot might be able to remain aligned with the formation only if the consensus position is within a certain region or the consensus velocity is not too high.
In all these cases, what one would like to add is a state constraint to the consensus problem.

Consensus problems with constraints have been studied from different perspectives in the literature.
A significant group of papers deals with the use of state projections on convex sets, mostly in discrete-time consensus problems and motivated by optimization algorithms \cite{Nedic2010Constrained}.
Projection-based methods for state constraint satisfaction have been introduced also for continuous-time consensus problems, using projection operators inspired by the adaptive control literature \cite{Wang2015Projection}, or logarithmic barrier functions \cite{Lee2011Constrained}. Continuous  flows can be used to solve convex intersection computation problems when the states of the nodes are not necessarily satisfying the constraints  for all time \cite{Shi2013TAC}.
Alternative approaches for imposing state constraints on consensus problems are proposed in e.g. \cite{Cao2015Consensus,Meng2017Consensus,Sun2013Consensus}.
A different situation of consensus with state constraints is the positive consensus problem studied in \cite{Valcher2016New}.
In this case, the aim is to achieve consensus while respecting the positivity of the state variables, representing e.g. quantities that are intrinsically nonnegative (masses, concentrations, etc.).

Other types of constrained consensus problems that have been considered in the literature include for instance the discarded consensus algorithm of \cite{LiuC12Discarded}, that discards the {state} of neighbors if they are outside of certain bounds, or the distributed averaging with flow constraints considered in \cite{Baric2011Distributed}.
Sometimes instead of state constraints one is interested in models with inputs constraints, representing e.g. actuator saturations, see e.g. \cite{Su2013Semiglobal,Li2011Consensus}.
The opinion dynamics literature offers several other contexts in which models are endowed with state constraints in order to better represent a phenomenon.
In \cite{bib:ManfrediAngeli2016} for instance, interactions are unilateral, i.e., are considered only if the state of the neighboring nodes is higher than the agent's state for optimistic models, or lower for pessimistic models.
A different approach, used in opinion dynamics, is proposed in the so-called bounded confidence models \cite{bib:BlondelHendrickxTsitsiklis2010,bib:BlondelHendrickxTsitsiklis2009}, in which states that are more distant than a certain threshold ignore each other.
The result is that these models produce clusters of opinions, and a local consensus value within each cluster.
Various variants of this opinion dynamics problem have been proposed, to accommodate other constraints in addition to bounded confidence. For instance in \cite{bib:LindmarkAll2016} the sign of the initial conditions is maintained throughout the opinion clustering process.

The problem we intend to study in this paper is different from all the aforementioned state-constrained consensus problems.
The main idea we want to introduce in a consensus problem is that we want to give to each agent the possibility of limiting the interval of values in which a consensus value can be accepted, and therefore force the agreed consensus value to belong to the intersection of all such intervals, if such intersection is nonempty.
The constraints we want to impose are however not classical hard constraints on the state variables.
Rather, they should only condition the range in which the steady state consensus value belongs to, but should be trespassable during the transient evolution.
To distinguish our problem from these other forms of consensus with hard-wired constraints, we call it {\em interval consensus}.

It is worth observing that our interval consensus problem is not related to the notion of ``bipartite interval consensus'' introduced in \cite{Meng2016Interval}.
In that paper, in fact, lack of strong connectivity of the graph is used to achieve some form of containment control (or leader-follower scheme \cite{bib:HongHuGao2006}), but no common value (monopartite or bipartite) is achieved.
In our problem, instead, the objective of the agents is to achieve a common consensus value, in spite of the interval constraints imposed by each of them.

Technically an agent implements an interval consensus by transmitting a value of its state which is saturated between an upper and a lower bound.
By limiting the transmitted state we can skip the projection step, and obtain the same result of imposing constraints on the consensus value although only asymptotically.
Practically it means that the agents keep seeking a compromise value fitting all constraints, and it is only through ``stubbornly'' transmitting a saturated value to its neighbors that an agent manages to carry the common consensus value within the interval imposed by its constraints.
Clearly, properties like the presence of a conserved quantity in average consensus, or the ``diffusion-like'' structure of any linear consensus algorithm are lost when the constraints become active.
In particular, when this happens the terms in the vector field that drives the consensus may no longer represent relative distances between agents states, meaning that the overall dynamical system behaves like a (Lipschitz continuous) switching system.
Nevertheless, in the paper we show that when the intersection of the intervals admissible by the agents is nonempty, a consensus is always achieved, and convergence must necessarily be to a value in the intersection.
In terms of the model, this translates into a system which is marginally stable inside the intersection of the allowed consensus intervals, but that is asymptotically stable outside it, because of the saturations.

In the paper we treat both the continuous-time and discrete-time interval consensus problems.
In both cases we assume that the graph of interactions is directed and strongly connected.
Needless to say, our interval consensus protocol respects the fully distributed nature of the problem, including for what concerns the individual upper and lower bounds, which are unknown to the other agents. While for the case of nonempty intersection of the admissible intervals (the most interesting from an application perspective) our results are complete, when the intersection is empty the analysis turns out to be more challenging, and we could obtain only partial results on the uniqueness and stability character of the equilibrium points.

A preliminary version of this paper appears in the conference proceedings of CDC 2017  \cite{FoShHuAl17}.
This conference paper concentrates exclusively on the nonempty interval intersection case.
All the material on the empty interval intersection case is presented here for the first time.

%%%%%%%%%%%%%%%%%%%%%%%%%%%%%%%%%%%%%%%%%%%%%%

\section{Problem Definition}
\subsection{The Model}
We consider a network with $n$ nodes indexed in the set $\mathrm{V}=\{1,\dots,n\}$. The structure of node interconnections is described by a simple directed graph $\mathrm{G}=(\mathrm{V},\mathrm{E})$, where each element in $\mathrm{E}$ is an ordered pair of two distinct
nodes in the set $\mathrm{V}$. The neighbor set of node $i$ in the graph $\mathrm{G}$ is denoted $\mathrm{N}_i=:\{j:(j,i)\in \mathrm{E}\}$.  Each edge $(j,i)\in\mathrm{E}$ is associated with a weight $a_{ij}>0$.

Each node $m$ holds a state $\mathbf{x}_m(t)\in\mathbb{R}$ at time $t\geq 0$. Instead of $ \mathbf{x}_m (t)$, the node transmits to its neighbors in $\mathrm{V}$ a value $\psi_m(\mathbf{x}_m(t))$ lying within an interval $\mathcal{I}_m:=[p_m,q_m]$, where
\begin{equation}
\psi_m(z) = \begin{cases}
p_m, & \text{ if   } z < p_m; \\
z, & \text{ if  } p_m \leq z \leq q_m; \\
q_m, & \text{ if  } z >q_m.
\end{cases}
\label{eq:psi}
\end{equation}
The evolution of $\mathbf{x}_i(t)\in\mathbb{R}$ is therefore described by
\begin{align}\label{eq:system1}
\frac{d}{dt}\mathbf{x}_i(t)= \sum_{j\in\mathrm{N}_i}a_{ij}\Big(\psi_j\big(\mathbf{x}_j(t) \big)-\mathbf{x}_i(t) \Big),\quad  i\in\mathrm{V}.
\end{align}
The nonlinear consensus system \eqref{eq:system1} will be studied in this paper.

\subsection{Examples}

A few more specific examples in which our notion of interval consensus is of interest are the following.

\begin{itemize}
\item \noindent{\em Achieving a price agreement among shareholders.}
Assume the board members of a company are negotiating a buy or sell order, and have to find an agreement among themselves on a price, price for which each of them is imposing boundaries.
If unanimity of the board is required, then the request of a consensus value that respects everybody's constraints has priority over for instance a consensus value which preserves the average of the initial bids.

\item \noindent{\em Load sharing under load assignment constraints.}
A network of computational units must share in equal parts a certain workload, under the constraint that each unit can allocate to the workload only a certain amount of resources, not known a priori to the other units. When is it possible for the units to agree on an equal load sharing policy and how?

\item \noindent{\em Social interactions under observer effect.}
The observer effect is a generalization of the DeGroot type social interaction rule \cite{DeGroot1974Reaching}, accounting for the fact that in face-to-face interactions opinions exchanged tend to be more ``moderate'' than they are in reality \cite{LeCompte1982Problems,Spano2005Potential}.
In particular, an agent tends to avoid assuming extremist opinions in a debate, but instead let them fall in a ``comfort interval'' shared with the other agents.
Seeking a consensus under such observer effect can be modeled as a saturation in the values of the transmitted opinions, as we do here.

\end{itemize}

In each of these cases, constraints are part of the problem, and if a consensus solution exists, then it has to respect them. There is however no need to impose that the transient dynamics respects them, i.e., the constraints are soft, not hard, as captured by the model \eqref{eq:system1}.

\subsection{Paper Outline}
The behavior of \eqref{eq:system1} depends crucially on the intersection of intervals
$\mcap_{m=1}^n \mathcal{I}_m $:
\begin{enumerate}
\item[(I):] When the intersection is nonempty, $\mcap_{m=1}^n \mathcal{I}_m \neq \emptyset $, then the system \eqref{eq:system1} always achieves a consensus value belonging to that intersection. This case is the most interesting from an application point of view. A complete analysis of its behavior is provided in both continuous-time (Section~\ref{sec:nonempty}) and discrete-time (Section~\ref{sec:DT:nonempty}).

\item[(II):] When instead the intersection is empty, $\mcap_{m=1}^n \mathcal{I}_m = \emptyset $,  then (at least) an equilibrium is always present, but it is typically not a consensus value.
As shown in Section~\ref{sec:empty}, only in some special cases uniqueness and asymptotic stability can be proven explicitly, although numerical simulations (Section~\ref{sec:examples}) suggest that a unique global attractor should be present in all cases.
\end{enumerate}

\section{Background Material}

Due to the nonlinearity in the network dynamics \eqref{eq:system1}, our work relies heavily on tools from nonlinear systems, non-smooth analysis, and robust consensus which are now briefly reviewed. 
\subsection{Monotonicity}\label{subsection-monotonicity}
Let $\mathbf{y}=(\mathbf{y}_1 \dots \mathbf{y}_n)^\top,\mathbf{z}=(\mathbf{z}_1 \dots \mathbf{z}_n)^\top\in\R^n$. We say $\mathbf{y}	\preceq \mathbf{z}$ if $\mathbf{y}_i\leq \mathbf{z}_i$ for all $i$. We next consider an autonomous dynamical system described by
\begin{align}\label{sys:monotone}
\frac{d}{dt}\mathbf{x}(t)=f(\mathbf{x}(t))=\Big(f_1(\mathbf{x}(t)) \dots f_n(\mathbf{x}(t) )\Big)^\top,
\end{align}
where $f(\cdot):\R^n \mapsto \R^n$ is Lipschitz continuous everywhere. Let $\bm{\phi}_t(\mathbf{y})$ be the solution of the system (\ref{sys:monotone}) with $\mathbf{x}(0)=\mathbf{y}$. We recall the following definition.
\begin{definition}
The system (\ref{sys:monotone}) is monotone if $\mathbf{y}\preceq\mathbf{z}$ implies $\bm{\phi}_t(\mathbf{y})\preceq\bm{\phi}_t(\mathbf{z})$ for all $\mathbf{y},\mathbf{z}\in \R^n$.
\end{definition}

An effective test for monotonicity of the dynamical systems from properties of the vector field relies on the so-called Kamke condition (pp.~581, Theorem 12.11, \cite{blanchini2008set}). The system \eqref{sys:monotone} is monotone if and only if
\begin{align*}
\mathbf{y}\preceq\mathbf{z}\ \mbox{and}\ \mathbf{y}_i=\mathbf{z}_i \quad \Longrightarrow \quad f_i(\mathbf{y})\leq f_i(\mathbf{z})
\end{align*}
holds for any $i=1,\dots,n$. It is easy to verify this condition for the network dynamics (\ref{eq:system1}). Therefore (\ref{eq:system1}) is a monotone dynamical system.

%---------------------------------------------------------------------------------

\subsection{Limit Set, Dini Derivatives, and Invariance Principle}
Consider the autonomous system \eqref{sys:monotone},
where $f:\R^d\to \R^d$ is a  continuous function. Then $\Omega_0\subset \R^d$ is called a {\it positively invariant
set} of \eqref{sys:monotone} if, for any $t_0\in\R$ and any $\mathbf{x}(t_0)\in\Omega_0$,
we have $\mathbf{x}(t)\in\Omega_0$, $t\geq t_0$, along  every solution $\mathbf{x}(t)$ of \eqref{sys:monotone}.

Let $\mathbf{x}:(\alpha,\omega)\rightarrow \R$ be a non-continuable  solution of
\eqref{sys:monotone} with initial condition $\mathbf{x}(0)=\mathbf{x}^0$, where $-\infty\leq \alpha < \omega\leq \infty$. We call $\mathbf{y}$ a  $\omega$-limit point of $\mathbf{x}(t)$ if there exists a  sequence $\{t_k\}$ with $\lim_{k\to \infty}t_k=\infty$ such that
$
\lim_{k\to \infty}\mathbf{x}(t_k)=\mathbf{y}.
$
The set of all $\omega$-limit points of $\mathbf{x}(t)$ is called the  $\omega$-limit set of  $\mathbf{x}(t)$,  and is denoted as $\Lambda^+(\mathbf{x}^0)$.
The following lemma is well-known  \cite{rou}.
\begin{lemma}\label{leminvariant}
Let  $\mathbf{x}(t)$ be a solution of \eqref{sys:monotone}. If $\mathbf{x}(t)$ is bounded, then   $\Lambda^+(\mathbf{x}^0)$ is nonempty, compact,  connected, and positively  invariant. Moreover,  there holds
$\mathbf{x}(t)\rightarrow \Lambda^+(\mathbf{x}^0)$ as $t \to \omega$ with $\omega=\infty$.
\end{lemma}

The upper {\it Dini derivative} of a continuous function $h: (a,b)\to \R$ ($-\infty\leq a<b\leq \infty$) at $t$ is defined as
\begin{equation*}
d^+h(t)=\limsup_{s\to 0^+}\, \frac{h(t+s)-h(t)}{s}.
\end{equation*}
When $h$ is continuous on $(a,b)$, $h$ is
non-increasing on $(a,b)$ if and only if $ d^+h(t)\leq 0$ for any
$t\in (a,b)$.

Now let $\mathbf{x}(t)$ be a solution of \eqref{sys:monotone} and let  $V:\R^d \to \R$ be a continuous, locally Lipschitz function. The Dini derivative of $V(\mathbf{x}(t))$, $d^+V(\mathbf{x}(t))$, thereby follows the above definition. On the other hand, one can also define
\begin{align}
d^+_f V(\mathbf{x})=\limsup_{s\to 0^+} \,\frac{V(\mathbf{x}+s f(\mathbf{x})) -V(\mathbf{x})}{s},
\end{align}
namely the upper Dini derivative of $V$ along the vector field \eqref{sys:monotone}. There holds that \cite{rou}
\begin{align}\label{guodong-2}
d^+_fV(\mathbf{x})\big|_{\mathbf{x}^\ast}=d^+V(\mathbf{x}(t))\big|_{t_\ast}
\end{align}
when putting $\mathbf{x}(t_\ast)=\mathbf{x}^\ast$. The next result is convenient for the calculation of the Dini derivative \cite{dan,Lin2007State}.

\begin{lemma}
\label{lemdini}
Let $V_i(\mathbf{x}): \R^d \to \R\;(i=1,\dots,n)$ be
$C^1$ and $V(\mathbf{x})=\max_{i=1,\dots,n}V_i(\mathbf{x})$. Let $\mathbf{x}(t) \in \R^d$ be an absolute continuous function over an interval $(a,b)$. If $
\mathsf{I}(t)=\{i\in \{1,2,\dots,n\}\,:\,V(\mathbf{x}(t))=V_i( \mathbf{x}(t))\}$
is the set of indices where the maximum is reached at $t$, then
$
d^+V(\mathbf{x}(t))=\max_{i\in\mathsf{ I}(t)}\dot{V}_i(\mathbf{x}(t)), t\in(a,b).
$
\end{lemma}

The following is the well-known LaSalle invariance principle.

\begin{lemma}[LaSalle (1968), Theorem 3.2 in \cite{rou}]
\label{lemma-LaSalle}
Let $\mathbf{x}(t)$ be a solution of \eqref{sys:monotone}. Let $V:\R^d \rightarrow \R$ be a continuous, locally Lipschitz function with $d^+V(\mathbf{x}(t))\leq 0$ on $[0,\omega)$. Then $\Lambda^+(\mathbf{x}^0)$ is contained in the union of all solutions that remain in $\mathcal{Z}:=\{\mathbf{x}:d^+_fV(\mathbf{x})=0\}$ on their maximal intervals of definition.
\end{lemma}

%---------------------------------------------------------------------------------

\subsection{Robust Consensus}
The following lemma deals with a robust version of the usual consensus problem, and it is a special case of Theorem 4.1 and Proposition 4.10  in \cite{shisiam}.
\begin{lemma}
\label{lemrobust}
Consider the following network dynamics defined over the digraph $\mathrm{G}$:
\begin{align}
	\frac{d}{dt}\mathbf{x}_i(t)=\sum_{j\in\mathrm{N}_i}a_{ij}\big(\mathbf{x}_j(t)-\mathbf{x}_i(t)\big)+w_i(t), \ i=1,\dots,n
\end{align}
where $w_i(t)$ is a piecewise continuous function. Let $\mathrm{G}$ contain a directed spanning tree. Denote  $ \|w(t)\|_{\infty}:= \max_{i \in\mathrm{V}} \sup_{t\in[0,\infty)} |w_i(t)|$.
Then for any $\epsilon>0$, there exists $\delta>0$ such that
\begin{equation*}
\|w(t)\|_{\infty}\leq \delta \quad  \Longrightarrow \quad \limsup_{t\rightarrow +\infty} \max_{i,j\in\mathrm{V}}\big |\mathbf{x}_i(t)-\mathbf{x}_j(t)\big|\leq  \epsilon
\end{equation*}
for all initial value $\mathbf{x}^0$.\end{lemma}

%==================================================================

\section{Nonempty Interval Intersection: Interval Consensus}
\label{sec:nonempty}
Denote  $\mathbf{x}(t)=(\mathbf{x}_1(t) \dots \mathbf{x}_n(t))^\top$ as the network state. Let $\mathbf{x}^0=(\mathbf{x}_1(0) \dots \mathbf{x}_n(0))^\top$ be the network initial value. The following theorem suggests that node state consensus can be enforced by the interval constraints node dynamics if the intervals admit some nonempty intersection.

\begin{theorem}
	\label{theorem:continuous}
Suppose $\mcap_{m=1}^n \mathcal{I}_m \neq \emptyset$ and let the underlying graph  $\mathrm{G}$ be strongly connected.  Then along the system \eqref{eq:system1}, for any initial value $\mathbf{x}^0$, there is $c^\ast(\mathbf{x}^0) \in \mcap_{m=1}^n \mathcal{I}_m  $ such that $$\lim_{t\to\infty} \mathbf{x}_i(t) =c^\ast, \ i\in\mathrm{V}.
$$
\end{theorem}
The condition $\mcap_{m=1}^n \mathcal{I}_m \neq \emptyset$ is equivalent to $p_\ast \leq q_\ast$ with
$$
p_\ast =\max_{i\in\mathrm{V}} p_i, \quad q_\ast=\min_{i\in\mathrm{V}} q_i.
$$
When such condition holds we have $\mcap_{m=1}^n \mathcal{I}_m=[p_\ast,q_\ast]$.

% Due to the nonlinearity in the network dynamics, the proof of Theorem \ref{theorem:continuous} relies heavily on some background knowledge in nonlinear systems, non-smooth analysis, and robust consensus.

\subsection{Proof of Theorem \ref{theorem:continuous}}
 We proceed in steps.

\medskip

\noindent {\bf Step 1}. Introduce $H(\mathbf{x}(t))=\max\big\{ \max_{i\in\mathrm{V}}\mathbf{x}_i(t), q_\ast \big\}$. Clearly $H$ is continuous and locally Lipschitz. If $H(\mathbf{x}(t))>q_\ast$, then $ \max_{i\in\mathrm{V}}\mathbf{x}_i(t)>q_\ast$ for $[t,t+\epsilon)$ for some sufficiently small $\epsilon$. Let $\mathsf{I}_0(t):=\{j: \mathbf{x}_j(t)=\max_{i\in\mathrm{V}}\mathbf{x}_i(t)\}$. As a result, from Lemma \ref{lemdini},
\begin{align}
d^+H(\mathbf{x}(t))
&=d^+\max_{i\in\mathrm{V}}\mathbf{x}_i(t)\nonumber\\
&=\max_{i\in \mathsf{I}_0(t)} \dot{\mathbf{x}}_i(t)\nonumber\\
&=\max_{i\in \mathsf{I}_0(t)}\Big[ -\sum_{j\in\mathrm{N}_i}a_{ij} (\mathbf{x}_i(t) -\psi_j(\mathbf{x}_j(t)))\Big].
\label{guodong-1}
\end{align}
Let $i_0\in \mathsf{I}_0(t)$. Then $\mathbf{x}_{i_0}(t)\geq \mathbf{x}_j(t)$ for all $j$. Moreover, by definition we have $q_j\geq q_\ast$, which implies:
\begin{itemize}
\item[(i).] $\psi_j(\mathbf{x}_j(t))\leq \mathbf{x}_j(t)$ if $\mathbf{x}_j(t) >q_\ast$;

\item[(ii).] $\psi_j(\mathbf{x}_j(t))\leq q_\ast$ if $\mathbf{x}_j(t) \leq q_\ast$.
\end{itemize}
Combining the two cases we can conclude that $\mathbf{x}_{i_0}(t) -\psi_j(\mathbf{x}_j(t))\geq 0$ since $\mathbf{x}_{i_0}(t)>q_\ast$. From (\ref{guodong-1}) we further know that
$d^+ H(\mathbf{x}(t))\leq 0$ if $H(\mathbf{x}(t))>q_\ast$. This in fact further assures that  if $H(\mathbf{x}(t_\ast))= q_\ast$, then $H(\mathbf{x}(t))=q_\ast$ for all $t\geq t_\ast$. We have proved that $H(\mathbf{x}(t))$ is a non-increasing function for all $t$.

Also introduce  $h(\mathbf{x}(t))=\min\big\{ \min_i\mathbf{x}_i(t), p_\ast \big\}$. The same argument leads to $d^+h(\mathbf{x}(t))\geq 0$, i.e., $h(\mathbf{x}(t))$ is a non-decreasing function for all $t$. Consequently, for  $V(\mathbf{x}(t))=H(\mathbf{x}(t))-h(\mathbf{x}(t))$, there holds $d^+V(\mathbf{x}(t))\leq 0$.

\medskip

\noindent {\bf Step 2}. Denote\footnote{More precisely, it is the Dini derivative of $V$ along system (\ref{eq:system1}). But by (\ref{guodong-2}), there is no harm writing it in this way.} $\mathcal{Z}:=\{\mathbf{x}: d^+V(\mathbf{x})=0\}$. In this step, we show $\mathcal{Z} \subseteq [p_\ast,q_\ast]^n$ when  $\mathrm{G}$ is strongly connected.

We use a contradiction argument. Let $\mathbf{x}^\ast=(\mathbf{x}_1^\ast \dots \mathbf{x}_n^\ast)^{T} \in \mathcal{Z}$ with $\mathbf{x}_\ast \notin [p_\ast,q_\ast]^n$. Then there must be a node $i$ satisfying $\mathbf{x}_{i}^\ast\notin [p_\ast,q_\ast]$. By symmetry we assume $\mathbf{x}_i^\ast >q_\ast$, and without loss of generality we let $\mathbf{x}_i^\ast= \max_{j\in\mathrm{V}} \mathbf{x}_j^\ast$. Let us   consider a solution $\mathbf{x}(t)$ of (\ref{eq:system1}) with $\mathbf{x}(0)=\mathbf{x}^\ast$.

Denote $\mathsf{I}_\ast:=\{j:\mathbf{x}_j^\ast =\mathbf{x}_i^\ast= \max_{k\in\mathrm{V}} \mathbf{x}_k^\ast \}$. Because $\mathrm{G}$ is strongly connected, along the system (\ref{eq:system1}), nodes in $\mathsf{I}_\ast$ will either be attracted by other nodes (if any) in $\mathrm{V}\setminus \mathsf{I}_\ast$ which hold values strictly smaller than $\mathbf{x}_i^\ast$, or simply  by $q_\ast$. Therefore, there is $\epsilon>0$ such that $\mathbf{x}_j(\epsilon)<\mathbf{x}_i^\ast $ for all $j$. This is to say, $H(\mathbf{x}(\epsilon))<H(\mathbf{x}(0))$ and therefore the trajectory cannot be within $\mathcal{Z}$. We have proved $\mathcal{Z} \subseteq [p_\ast,q_\ast]^n$.

Now by Lemma \ref{lemma-LaSalle}, $\Lambda^+(\mathbf{x}^0)$ is always contained in $\mathcal{Z}$, and therefore $\Lambda^+(\mathbf{x}^0)\subseteq [p_\ast,q_\ast]^n$. Further by Lemma~\ref{leminvariant}, there holds\footnote{From the property of $V$, each trajectory is obvious contained in a compact set with $\omega=\infty$.}
\begin{align}\label{guodong-3}
\mathbf{x}(t) \rightarrow [p_\ast,q_\ast]^n
\end{align}
as $t\rightarrow \infty$.

\medskip

\noindent{\bf Step 3.} By (\ref{guodong-3}), for any $\delta>0$, there is $T(\delta)>0$ such that along (\ref{eq:system1}), there holds
$$
\big|\mathbf{x}_i(t)-\psi_i(\mathbf{x}_i(t))\big|\leq \delta
$$
for all $t\geq T$. We can therefore rewrite (\ref{eq:system1}) as
\begin{align}
\frac{d}{dt}\mathbf{x}_i(t)=-\sum_{j\in\mathrm{N}_i}a_{ij}\big( \mathbf{x}_i(t)-\mathbf{x}_j(t)\big) +w_i(t),
\end{align}
with
$$
w_i(t):=\sum_{j\in\mathrm{N}_i} a_{ij}\big(\psi_j(\mathbf{x}_j(t))-\mathbf{x}_j(t)\big),
$$
and conclude that $|w_i(t)|\leq \delta$ for $t\geq T(\delta)$. Noticing that $\delta$ can be arbitrary and invoking Lemma \ref{lemrobust} for time interval $[T,\infty)$, we conclude that along (\ref{eq:system1}),
\begin{align}\label{guodong-4}
\limsup_{t\rightarrow +\infty} \max_{i,j\in\mathrm{V}}\big |\mathbf{x}_i(t)-\mathbf{x}_j(t)\big|=0.
\end{align}

\medskip

\noindent {\bf Step 4}. In this step, we finally show that each $\mathbf{x}_i(t)$ admits a finite limit. Let $c^\ast$ be a limit point of $\mathbf{x}_j(t)$ for a fixed $j$. Based on the fact that $\mathcal{Z} \subseteq [p_\ast,q_\ast]^n$, there must hold $c^\ast \in[p_\ast,q_\ast]$. If $p_\ast=q_\ast$, the result already holds. We assume $p_\ast<q_\ast$ in the following.

  According to (\ref{guodong-4}), for any $\epsilon>0$, there exists $t_\ast>0$ such that
\begin{align}
\big|\mathbf{x}_i(t_\ast) -c^\ast \big|\leq \epsilon.
\end{align}
for all $i$. There are three cases.
\begin{itemize}
\item[(i)] Let $c^\ast\in (p_\ast,q_\ast)$. We let $\epsilon$ be sufficiently small so that
$$
p_\ast <c^\ast -\epsilon \leq \mathbf{x}_i(t_\ast) \leq c^\ast+\epsilon <q_\ast
$$
for all $i$. This means the system (\ref{eq:system1}) is a standard consensus dynamics for $t\geq t_\ast$ because $\psi_i(\mathbf{x}_i(t))=\mathbf{x}_i(t)$ for all $t\geq t_\ast$. Of course all $\mathbf{x}_i(t)$ converge to the same limit, which must be $c^\ast$.

\item[(ii)] Let $c^\ast= q_\ast$. We let $\epsilon$ be sufficiently small so that
$$
 p_\ast <q_\ast-\epsilon <q_\ast.
$$
As a result,  $p_\ast<\mathbf{x}_i(t_\ast)\leq q_\ast+\epsilon$ for all $i$. Repeating the argument we used in Step 1, it is easy to see that
$$
\max_{i\in\mathrm{V}} \mathbf{x}_i(t)
$$
is non-increasing for $t\geq t_\ast$, and therefore it converges to a finite limit, say $M_\ast$. While from (\ref{guodong-4}), $\min_i \mathbf{x}_i(t)$ must converge to the same limit $M_\ast$. This leaves $c^\ast=q_\ast=M_\ast$ to be the only possibility, and all $\mathbf{x}_i(t)$ converge to $q_\ast$.

\item[(iii)] Let $c^\ast= p_\ast$. The argument is symmetric to Case (ii). By first showing that $\min_i \mathbf{x}_i(t)$ converges with $p_\ast<p_\ast+\epsilon <q_\ast$, we know all $\mathbf{x}_i(t)$ converge to $p_\ast$.
\end{itemize}
We have now proved that all $\mathbf{x}_i(t)$ will converge to a common limit within $[p_\ast,q_\ast]$. The proof is complete. \hfill$\square$

%=================================================================

\section{Empty Interval Intersection: Existence and Stability of Equilibria}
\label{sec:empty}
In this section, we study the network dynamics \eqref{eq:system1} when the intervals $\mathcal{I}_m$ admit an empty intersection. To this end, we denote $\mathbf{x}(t;\mathbf{y})$ as the solution of \eqref{eq:system1} with the initial condition $\mathbf{x}(0)=\mathbf{y}$.
Denote $\underline{p}=\min_{i\in \mathrm{V}} p_i$ and $\overline{q}=\max_{i\in \mathrm{V}} q_i$. It is obvious that $conv\left(\mcup_{m\in\mathrm{V}}\mathcal{I}_m\right)=[\underline{p},\overline{q}]$, where $conv$ denotes the convex hull of a set. It turns out that, regardless of the network topology $\mathrm{G}$ and the intervals $\mathcal{I}_m$, the nonlinearity of \eqref{eq:system1} always defines equilibria dynamics.

\begin{theorem}\label{theorem:equilibrium}
 The system \eqref{eq:system1} has at least one equilibrium. In fact, all equilibria of the system \eqref{eq:system1} lie within $[\underline{p},\overline{q}]^n$ if $\mathrm{G}$ is strongly connected.
\end{theorem}

Naturally we are interested in the stability of the equilibria. We introduce the following definitions.

\begin{definition}\label{definition:equilibriumclass}
 An equilibrium $\mathbf{e}=(\mathbf{e}_1\dots \mathbf{e}_n)^\top$ is an equi-unconstrained  equilibrium if $\mathbf{e}_m$ is an interior point of $[p_m,q_m]$ for all $m\in\mathrm{V}$; an equi-constrained equilibrium if $\mathbf{e}_m$ is an interior point of $\mathbb{R}\setminus[p_m,q_m]$ for all $m\in\mathrm{V}$.\end{definition}

\begin{definition}
(i) An equilibrium $\mathbf{e} $ is locally   stable if  for any $\delta>0$, there exists $\epsilon >0$ such that $\|\mathbf{x}(t;\mathbf{y})\|\leq \delta$ for all $t\geq 0$ and all $\|\mathbf{y}-\mathbf{e}\|\leq \epsilon$;

(ii) An equilibrium $\mathbf{e} $ is locally asymptotically  stable if  for any $\delta>0$, there exists $\epsilon >0$ such that $\|\mathbf{x}(t;\mathbf{y})\|\leq \delta$ for all $t\geq 0$ and all $\|\mathbf{y}-\mathbf{e}\|\leq \epsilon$, and  $\lim_{t\to \infty}\mathbf{x}(t;\mathbf{y})=\mathbf{e}$ for all $\|\mathbf{y}-\mathbf{e}\|\leq \epsilon$.
\end{definition}

    We present the following result for the stability of equi-unconstrained or equi-constrained equilibria.

\begin{theorem}\label{thm:local-stability}
 Then the following statements hold.

(i) Every equi-unconstrained equilibrium is locally stable;

(ii) Every equi-constrained equilibrium  $\mathbf{e}=(\mathbf{e}_1,\dots,\mathbf{e}_n)^\top$ of the system \eqref{eq:system1} is locally asymptotically stable if $\mathrm{N}_i\neq \emptyset$ for all $i\in\mathrm{V}$.
\end{theorem}

Apparently the classes of equi-unconstrained and equi-constrained equilibria only cover a fraction of possible equilibria of the network dynamics. With pairwise disjoint constraint intervals, i.e., $ \mathcal{I}_{m_1} \cap \mathcal{I}_{m_2} = \emptyset $ $ \forall \, m_1, \, m_2 \in \mathrm{V}$, we can establish a full picture regarding the stability of the equilibria.
\begin{theorem}\label{thm:disjoint-local-stability}
 Let the graph $\mathrm{G}$ be strongly connected and suppose the $\mathcal{I}_m$ are pairwise disjoint. Then for the system \eqref{eq:system1} the following statements hold.
 \begin{itemize}
 \item[(i)] There cannot exist   equi-unconstrained equilibria;
 \item[(ii)] Every equilibrium is locally asymptotically stable.
\end{itemize}
\end{theorem}

We conjecture that the system \eqref{eq:system1} should have a unique equilibrium which is globally attractive when the interaction graph $\mathrm{G}$ is strongly connected and  the $\mathcal{I}_m$ are pairwise disjoint. It seems that there are some major difficulties in establishing such an assertion  due to the nonlinear node dynamics. Nonetheless, we manage to prove the following result for directed graphs with the in-degree  no more than one at the majority of the nodes.
\begin{theorem}\label{thm:cycle-convergence}
Assume that $|\mathrm{N}_m|\leq 2$ for all $m\leq \mathrm{V}$ with the inequality holding at most for exactly one node. Suppose the $\mathcal{I}_m$ are pairwise disjoint.  Then along the system \eqref{eq:system1},  there exists  $\mathbf{d}^{^\ast}=(\mathbf{d}^{^\ast}_1 \dots \mathbf{d}^{^\ast}_n)^\top \in \R^n$ such that $$\lim_{t\to\infty} \mathbf{x}_i(t;\mathbf{x}^0) =\mathbf{d}_i^{^\ast}, \quad i\in\mathrm{V}
$$
for all initial value $\mathbf{x}^0$.
\end{theorem}

\begin{remark}
The underlying graph  $\mathrm{G}$ is termed a symmetric undirected graph if   $(i,j)\in\mathrm{V}$ if and only if $(j,i)\in\mathrm{V}$, and  $a_{ij}=a_{ji}$ for all $(i,j)\in\mathrm{V}$. Undirected graphs would not help too much to simplify the stability analysis because there can be the case with $\psi_i(\mathbf{x}_i)=\mathbf{x}_i$ while  $\psi_j(\mathbf{x}_j)=p_j$. Therefore locally the node interactions could be essentially directional even with bidirectional interactions.
\end{remark}

%------------------------------------------------------------------------------

\subsection{Proof of Theorem \ref{theorem:equilibrium}}
We rewrite the system \eqref{eq:system1} by
\begin{align}
\frac{d}{dt}\mathbf{x}(t)=g(\mathbf{x}(t))=\Big ( g_1(\mathbf{x}(t)) \dots g_n(\mathbf{x}(t))\Big)^\top
\end{align}
with $g_i(\mathbf{x}(t))=\sum_{j\in\mathrm{N}_i}a_{ij}\big(\psi_j\big(\mathbf{x}_j(t) \big)-\mathbf{x}_i(t) \big)$. Now let $\mathbf{x}^0\in[\underline{p},\overline{q}]^n$. Then it is straightforward to verify that $\mathbf{x}(t;\mathbf{x}^0) \in[\underline{p},\overline{q}]^n$ for all $t\geq 0$ because the vector field $g$ is pointing inwards the n-dimensional cube $[\underline{p},\overline{q}]^n$. This leads to the following  lemma.

\begin{lemma} The set $[\underline{p},\overline{q}]^n$ is positively invariant along the system (\ref{eq:system1}).
\end{lemma}

Therefore, $\mathbf{x}(t;\cdot)$ defines a continuous mapping from $[\underline{p},\overline{q}]^n$ to itself. By the famous Brouwer fixed-point theorem, there is at least one point $\mathbf{e} \in [\underline{p},\overline{q}]^n$ satisfying that $\mathbf{x}(t;\mathbf{e})=\mathbf{e}$, i.e., $\mathbf{e}$ is a fixed point. We have proved existence of equilibria of the network dynamics within the set $[\underline{p},\overline{q}]^n$. In order to further prove that there can be no equilibrium outside the set $[\underline{p},\overline{q}]^n$ when  $\mathrm{G}$ is strongly connected, we need the following lemma.

\begin{lemma}\label{lem:lemma-unioninterval}
Let the graph $\mathrm{G}$ be strongly connected. Then along the system (\ref{eq:system1}) there holds for any $\mathbf{x}^0\in\R^n$ that $$\lim_{t\to\infty}{\rm dist}\Big(\mathbf{x}(t),[\underline{p},\overline{q}]^n\Big)=0.
$$
\end{lemma}
{\it Proof.} Define $\beta(t)=\max_{m\in\mathrm{V}}\mathbf{x}_m(t)$ and again let $\mathsf{I}_0(t):=\{j: \mathbf{x}_j(t)=\max_{i\in\mathrm{V}}\mathbf{x}_i(t)\}$. Note that, $\psi_m\big(\mathbf{x}_m(t) \big)\leq \overline{q}$ for all $m\in\mathrm{V}$ and for all $t\geq 0$. From Lemma \ref{lemdini} and noticing the structure of the node dynamics, there holds that if $ \beta(t)\geq \overline{q}$, then \begin{align}
d^+\beta(t)& =\max_{i\in\mathsf{I}_0(t)} \frac{d}{dt} \mathbf{x}_i(t)\nonumber\\
&=\max_{i\in\mathsf{I}_0(t)}  \sum_{j\in\mathrm{N}_i}a_{ij}\big(\psi_j\big(\mathbf{x}_j(t) \big)-\mathbf{x}_i(t) \big)\nonumber\\
&\leq \max_{i\in\mathsf{I}_0(t)}  \sum_{j\in\mathrm{N}_i}a_{ij}\big(\overline{q}-\mathbf{x}_i(t) \big)\nonumber\\
&=\max_{i\in\mathsf{I}_0(t)}  \sum_{j\in\mathrm{N}_i}a_{ij}\big(\overline{q}-\beta(t) \big)\nonumber\\
&\leq \min\{a_{ij}>0:(j,i)\in\mathrm{E}\}\big(\overline{q}- \beta(t)\big)
\end{align}
when  $\mathrm{G}$ is strongly connected.
As a result, we can obtain $\limsup_{t\to\infty}\beta(t)\leq \overline{q}$. A symmetric argument leads to the fact that  $\liminf_{t\to\infty}\beta(t)\geq \overline{p}$. We have now proved the desired lemma. \hfill$\square$

Based on Lemma \ref{lem:lemma-unioninterval}, obviously every equilibrium must be within the set $[\underline{p},\overline{q}]^n$ when  $\mathrm{G}$ is strongly connected. This proves Theorem \ref{theorem:equilibrium}.

\subsection{Proof of Theorem \ref{thm:local-stability}}

\noindent (i) Let the equilibrium $\mathbf{e}$ be an equi-unconstrained equilibrium, i.e., $\mathbf{e}_m$ is an interior point of $[p_m,q_m]$ for all $m\in\mathrm{V}$.

%Denote $\bm{\gamma}(t)=(\bm{\gamma}_1(t) \dots$ $\bm{\gamma}_n(t))^\top$ with $\bm{\gamma}(t)=\mathbf{x}(t)-\mathbf{e}$.

Under the above condition on $\mathbf{e}$ there exists $\epsilon>0$ such that for
$$
\mathbb{B}_\epsilon(\mathbf{e}):=\{\mathbf{y}\in\mathbb{R}^n:\|\mathbf{y}-\mathbf{e}\|<\epsilon\}
$$ there holds
\begin{align}\label{100}
\frac{d}{dt}\mathbf{x}(t) =- \mathbf{L}\mathbf{x}(t), \ \mathbf{x}(t)\in\mathbb{B}_\epsilon(\mathbf{e})
\end{align}
where $\mathbf{L}$ is the network Laplacian.  Clearly (\ref{100}) is standard consensus dynamics. It is easy to show that $\mathbb{B}_\epsilon(\mathbf{e})$ is an invariant set of the network system (\ref{eq:system1}). Therefore $\mathbf{e}$ is locally stable.

\noindent (ii) Let the equilibrium $\mathbf{e}$ be an equi-constrained equilibrium, i.e., $\mathbf{e}_m$ is an interior point of $\R\setminus[p_m,q_m]$ for all $m\in\mathrm{V}$.
\begin{align}\label{101}
\frac{d}{dt}\big(\mathbf{x}(t)-\mathbf{e}\big) =- \mathbf{D}\big(\mathbf{x}(t)-\mathbf{e}\big), \ \mathbf{x}(t)\in\mathbb{B}_\epsilon(\mathbf{e})
\end{align}
where $\mathrm{D}={\rm diag}(\mathbf{d}_1\dots \mathbf{d}_n)$ is the degree matrix with $\mathbf{d}_i=\sum_{j\in\mathrm{N}_i}a_{ij}$. Now let $\mathbf{x}^0\in \mathbb{B}_\epsilon(\mathbf{e})$. Then there exists $\mu>0$ such that $\mathbf{x}(t;\mathbf{x}^0)\in\mathbb{B}_\epsilon(\mathbf{e}) $ for $t\in[0,\mu]$ simply by continuity of the trajectory. However, along the interval  $t\in[0,\mu]$  for the system (\ref{101}) there holds $\|\mathbf{x}(t;\mathbf{x}^0)- \mathbf{e}\|\leq \|\mathbf{x}^0- \mathbf{e}\|$. Therefore again we have shown that $\mathbb{B}_\epsilon(\mathbf{e})$ is an invariant set in such case along the network system (\ref{eq:system1}). The local asymptotical stability of $\mathbf{e}$ is then straightforward to verify.

We have now proved Theorem \ref{thm:local-stability}.

%-----------------------------------------------------------------------------------------------------

\subsection{Proof of Theorem \ref{thm:disjoint-local-stability}}
Without loss of generality we assume $p_1<p_2<\dots <p_n$ and therefore $q_1<q_2<\dots <q_n$.
We first establish a technical lemma strengthening the statement of Lemma \ref{lem:lemma-unioninterval}.
\begin{lemma}\label{lemma-finiteentering}
Let the graph $\mathrm{G}$ be strongly connected. Suppose the $\mathcal{I}_m$ are pairwise disjoint with $p_1<p_2<\dots <p_n$. Then along the system (\ref{eq:system1}) there holds that for any $\mathbf{x}^0\in\R^n$, there exists $T(\mathbf{x}^0)$ such that $$
\lim_{t\to\infty}\big\|\mathbf{x}(t)\big\|_{[q_1,p_n]^n}=0
$$
\end{lemma}
{\it Proof.} Since the $\mathcal{I}_m$ are pairwise disjoint with $p_1<p_2<\dots <p_n$, we have $\psi_j\big(\mathbf{x}_j(t) \big)\leq q_{n-1}$ for all $j=1,\dots,n-1$. Thus,
\begin{align}
\frac{d}{dt}\mathbf{x}_n(t)&=\sum_{j\in\mathrm{N}_n}a_{nj}\big(\psi_j\big(\mathbf{x}_j(t) \big)-\mathbf{x}_n(t) \big)\nonumber\\
&\leq\sum_{j\in\mathrm{N}_n}a_{nj}\big(q_{n-1}-\mathbf{x}_n(t) \big),
\end{align}
which implies that $\limsup_{t\to\infty}\mathbf{x}_n(t)\leq q_{n-1}$. Therefore, for any $\mathbf{x}^0$, there is $T_1(\mathbf{x}^0)>0$ that
\begin{align}
\psi_n\big(\mathbf{x}_n(t) \big)=p_n, \ t\geq T_1.
\end{align}

Let $\beta(t)$ and $\mathsf{I}_0(t)$ be defined as in the proof of   Lemma \ref{lem:lemma-unioninterval}. If $ \beta(t)\geq p_n$, then \begin{align}
d^+\beta(t)& =\max_{i\in\mathsf{I}_0(t)} \frac{d}{dt} \mathbf{x}_i(t)\nonumber\\
&=\max_{i\in\mathsf{I}_0(t)}  \sum_{j\in\mathrm{N}_i}a_{ij}\big(\psi_j\big(\mathbf{x}_j(t) \big)-\mathbf{x}_i(t) \big)\nonumber\\
&\leq \max_{i\in\mathsf{I}_0(t)}  \sum_{j\in\mathrm{N}_i}a_{ij}\big(p_n-\beta(t)\big)\nonumber\\
&\leq \min\{a_{ij}>0:(j,i)\in\mathrm{E}\}\big(p_n- \beta(t)\big)
\end{align}
when  $\mathrm{G}$ is strongly connected. This in turn leads to the fact that  $\limsup_{t\to\infty}\mathbf{x}_n(t)\leq p_{n}$.

A symmetric argument will give us  $\liminf_{t\to\infty}\mathbf{x}_n(t)\geq q_{1}$ based on the fact that there is $T_2(\mathbf{x}^0)>0$ that $
\psi_1\big(\mathbf{x}_1(t) \big)=q_1, \ t\geq T_2$. We have now completed the proof of the desired lemma. \hfill$\square$

We are now ready to prove   Theorem \ref{thm:disjoint-local-stability}. Let  $\mathbf{e}=(\mathbf{e}_1\dots \mathbf{e}_n)^\top$ be an equilibrium. From Lemma \ref{lemma-finiteentering} and its proof there must hold $\psi_{1}(\mathbf{e}_1)=q_1$ and $\psi_{n}(\mathbf{e}_n)=p_n$ with $\mathbf{e}_1>q_1$ and $\mathbf{e}_n<p_n$. We denote $\mathrm{V}^\dag$ as the node set with
$$
\mathrm{V}^\dag:=\big\{m\in\mathrm{G}:\ \mathbf{e}_m\in \R\setminus (p_{m},q_{m})\big\}.
$$
There holds $\mathrm{V}^\dag\neq \emptyset$ since $\{1,n\}\in\mathrm{V}^\dag$. Let us introduce
$$
\mathbb{N}_\epsilon^+(\mathbf{e}):=\big\{ \mathbf{y}=(\mathbf{y}_1 \dots \mathbf{y}_n)^\top: \mathbf{y}_i\in [\mathbf{e}_i,\mathbf{e}_i+\epsilon)\big\}.
$$
For sufficiently small $\epsilon$ the network dynamics can be rewritten as
\begin{align}
\frac{d}{dt}\big(\mathbf{x}(t)-\mathbf{e}\big) =- \mathbf{H}\big(\mathbf{x}(t)-\mathbf{e}\big), \ \mathbf{x}(t)\in\mathbb{N}_\epsilon^+(\mathbf{e})
\end{align}
where $\mathbf{H}$ depends on the structure of $\mathrm{V}^\dag$.

Now that $\mathrm{G}$ is strongly connected and $\{1,n\}\in\mathrm{V}^\dag$, the matrix $\mathbf{H}$ is Hurwitz since it is weakly diagonally dominant and irreducible with negative diagonal entries \cite{Taussky1949Recurring}. Therefore, given $I_n$ as the $n$ dimensional identity matrix, there exists a unique symmetric positive definite matrix $P$ such that
\begin{align}\label{lyapunov}
P\mathbf{H}+\mathbf{H}^\top P =-I_n.
\end{align}
We establish two facts.

\begin{itemize}
\item[F1.] $\mathbb{N}^+(\mathbf{e})=\big\{ \mathbf{y}=(\mathbf{y}_1 \dots \mathbf{y}_n)^\top: \mathbf{y}_i\geq \mathbf{e}_i\big\}$ is positively invariant along the network dynamics (\ref{eq:system1}). This is an immediate conclusion from the monotonicity of the system (\ref{eq:system1}) established in Subsection \ref{subsection-monotonicity}.

\item[F2.]  From the Lyapunov equation (\ref{lyapunov}) we can routinely obtain
$$
\|\mathbf{x}(t)-\mathbf{e}\|^2 \leq e^{-t/\lambda_{\rm max}(P)}\frac{\lambda_{\rm max}(P)}{\lambda_{\rm min}(P)} \|\mathbf{x}(0)-\mathbf{e}\|^2
$$
along the linear dynamics $\frac{d}{dt}\big(\mathbf{x}(t)-\mathbf{e}\big) =- \mathbf{H}\big(\mathbf{x}(t)-\mathbf{e}\big)$.
\end{itemize}

Combining the two facts, we conclude that there exists $\delta>0$ such that for any $\mathbf{x}^0\in \mathbb{N}^+_\delta(\mathbf{e})$, there holds $\mathbf{x}(t;\mathbf{x}^0)\in \mathbb{N}^+_\epsilon(\mathbf{e})$ for all $t\geq 0$. Further, it is straightforward to verify that
$\lim_{t\to\infty}\mathbf{x}(t;\mathbf{x}^0)=\mathbf{e}$ if $\mathbf{x}^0\in \mathbb{N}^+_\delta(\mathbf{e})$.

In order to complete the proof we also need to consider the other $2^{n}-1$ orthants around the equilibrium.
$$
\mathbb{N}_\epsilon^{o}(\mathbf{e}):=\big\{ \mathbf{y}=(\mathbf{y}_1 \dots \mathbf{y}_n)^\top: \mathbf{y}_i\in \mathcal{J}^o_i (\mathbf{e},\epsilon)\big\}
$$
with  $\mathcal{J}_i^o (\mathbf{e},\epsilon)\in \Big\{ [\mathbf{e}_i,\mathbf{e}_i+\epsilon), (\mathbf{e}_i-\epsilon,\mathbf{e}_i]  \Big\}$ for each $i$. For each $\mathbb{N}_\epsilon^{o}(\mathbf{e})$ we can use the transform
\begin{align}
\mathbf{y}_i=-\mathbf{x}_i, &\ {\rm if}\ \mathcal{J}_i^o (\mathbf{e},\epsilon)=(\mathbf{e}_i-\epsilon,\mathbf{e}_i];\nonumber\\
\mathbf{y}_i=\mathbf{x}_i, &\ {\rm if}\ \mathcal{J}_i^o (\mathbf{e},\epsilon)=[\mathbf{e}_i,\mathbf{e}_i+\epsilon).
\end{align}
Then the problem will become identical to the case of $\mathbb{N}^+_\delta(\mathbf{e})$. This concluded the proof of  Theorem \ref{thm:disjoint-local-stability}.
\subsection{Proof of Theorem \ref{thm:cycle-convergence}}

The proof relies on the intermediate statements in the proof of Lemma \ref{lemma-finiteentering} that
$$
\psi_1\big(\mathbf{x}_1(t) \big)=q_1, \ \psi_n\big(\mathbf{x}_n(t) \big)=p_n
$$
for all $t\geq \max\{T_1,T_2\}$.

Since $|\mathrm{N}_m|\leq 2$ for all $m\in \mathrm{V}$ with the inequality holding at most for exactly one node and from the strong connectivity of $\mathrm{G}$, there must be  a node  $i_0\notin\{1,n\}$ satisfying that $\mathrm{N}_{i_0}=\{1\}$, $\mathrm{N}_{i_0}=\{n\}$, or $\mathrm{N}_{i_0}=\{1,n\}$. Consequently, from the network dynamics (\ref{eq:system1}) it is obvious that there is $\mathbf{d}^{^\ast}_{i_0}$ such that
$
\lim_{t\to \infty}\mathbf{x}_{i_0}(t)=\mathbf{d}^{^\ast}_{i_0}.
$ The continuity of $\psi_{i_0}(\cdot)$ in turn implies
 $$
\lim_{t\to\infty}\psi_{i_0}(\mathbf{x}_{i_0}(t))=\psi_{i_0}(\mathbf{d}^{^\ast}_{i_0}).
$$
Next, there exists a node $i_1 \neq i_0$ with $\mathrm{N}_{i_1}\in\{i_0,1,n\}$ since $\mathrm{G}$ is strongly connected. From the facts that  $\psi_{m}(\mathbf{x}_m(t))$ converges to finite limits for $m\in\{i_0,1,n\}$, we further know that there is $\mathbf{d}^{^\ast}_{i_1}$ such that
$
\lim_{t\to \infty}\mathbf{x}_{i_1}(t)=\mathbf{d}^{^\ast}_{i_1}.
$ This recursion can be proceeded until all nodes in the set $\mathrm{V}$ have been visited, which implies the conclusion of Theorem \ref{thm:cycle-convergence}.

%=================================================================

\section{Nonempty Interval Intersection in Discrete Time}
\label{sec:DT:nonempty}
Let us consider the discrete-time network dynamics analogous  to (\ref{eq:system1}) as below:
\begin{align}\label{eq:system-discrete}
\mathbf{x}_i(t+1)&= \mathbf{x}_i(t)+\epsilon \sum_{j\in\mathrm{N}_i}a_{ij}\Big(\psi_j\big(\mathbf{x}_j(t) \big)-\mathbf{x}_i(t) \Big)\nonumber\\
&= \Big(1-\epsilon \sum_{j\in\mathrm{N}_i}a_{ij} \Big)\mathbf{x}_i(t)+ \epsilon \sum_{j\in\mathrm{N}_i}a_{ij}\psi_j\big(\mathbf{x}_j(t)\big)
\end{align}
for all $i\in\mathrm{V}.$
Clearly (\ref{eq:system-discrete}) is the Euler approximation of (\ref{eq:system1}) with $\epsilon$ be a small step size.

\begin{theorem}
	\label{theorem:discrete}
Suppose $\mcap_{m=1}^n \mathcal{I}_m \neq \emptyset$ and let the underlying graph  $\mathrm{G}$ be strongly connected. Suppose  $
\epsilon <1/\max_{i\in\mathrm{V}} \sum_{j\in\mathrm{N}_i}a_{ij}$. Then along the system \eqref{eq:system-discrete}, for any initial value $\mathbf{x}^0$, there is $c^\ast(\mathbf{x}^0) \in \mcap_{m=1}^n \mathcal{I}_m  $ such that $$\lim_{t\to\infty} \mathbf{x}_i(t) =c^\ast, \ i\in\mathrm{V}.
$$
\end{theorem}
{\it Proof}. The proof has to rely on some new development from  the proof of Theorem \ref{theorem:continuous} since  we cannot use LaSalle invariance principle. We continue to use the definitions of $H(\mathbf{x}(t))$, $h(\mathbf{x}(t))$, and $V(\mathbf{x}(t))$, but defined over the discrete-time system (\ref{eq:system-discrete}). Again we proceed in steps.

\medskip

\noindent {\bf Step 1}. In this step, let us establish the monotonicity of the    functions $H(\mathbf{x}(t))$ and $h(\mathbf{x}(t))$. We introduce a function $\mathbf{I}_{a}^+(\cdot)$ by $\mathbf{I}_{a}^+(y)=y,y>a$ and $\mathbf{I}_{a}^+(y)=a,y\leq a$.
Therefore,
\begin{align}
&H(\mathbf{x}(t+1))
=\mathbf{I}_{q_{\ast}}^+ \bigg( \max_{i\in\mathrm{V}}  \mathbf{x}_i(t+1) \bigg)\nonumber\\
&=\mathbf{I}_{q_{\ast}}^+ \bigg( \max_{i\in\mathrm{V}}   \Big((1-\epsilon \sum_{j\in\mathrm{N}_i} a_{ij})\mathbf{x}_i(t)+ \epsilon  \sum_{j\in\mathrm{N}_i} a_{ij}\psi_j (\mathbf{x}_j(t))\Big) \bigg)\nonumber\\
&\leq \mathbf{I}_{q_{\ast}}^+ \bigg( \max_{i\in\mathrm{V}}  \Big((1-\epsilon \sum_{j\in\mathrm{N}_i}a_{ij})\mathbf{I}_{q_{\ast}}^+(\mathbf{x}_i(t))+ \epsilon \sum_{j\in\mathrm{N}_i}a_{ij} \mathbf{I}_{q_{\ast}}^+(\mathbf{x}_j(t))\Big) \bigg)\nonumber\\
&\leq \mathbf{I}_{q_{\ast}}^+ \bigg( \max_{i\in\mathrm{V}} \Big((1-\epsilon \sum_{j\in\mathrm{N}_i} a_{ij})\mathbf{I}_{q_{\ast}}^+(\max_{j\in\mathrm{V}} \mathbf{x}_j(t))\nonumber\\
&+\epsilon\sum_{j\in\mathrm{N}_i} a_{ij} \mathbf{I}_{q_{\ast}}^+(\max_{j\in\mathrm{V}} \mathbf{x}_j(t))\Big) \bigg)\nonumber\\
&= \mathbf{I}_{q_{\ast}}^+\Big(  \mathbf{I}_{q_{\ast}}^+(\max_{j\in\mathrm{V}} \mathbf{x}_j(t)) \Big)\nonumber\\
&= H(\mathbf{x}(t)),
\end{align}
where the first inequality holds due to the definition of $\mathbf{I}_{q_{\ast}}^+(\cdot)$, and the second inequality is based on  the monotonicity of $\mathbf{I}_{q_{\ast}}^+(\cdot)$ as well as the assumption that $\epsilon<1/\max_{i\in\mathrm{V}} \sum_{j\in\mathrm{N}_i} a_{ij}$. We can use a symmetric argument to establish  $h(\mathbf{x}(t+1))\geq h(\mathbf{x}(t))$.

\medskip

\noindent {\bf Step 2}. From the conclusion of the previous analysis, there are two constants $H_\ast$ and $h_\ast$ such that
\begin{equation*}
\lim_{t\to \infty } H(\mathbf{x}(t+1))=H_\ast,\ \lim_{t\to \infty } h(\mathbf{x}(t+1))=h_\ast.
\end{equation*}
Note that there always holds  $H_\ast \geq q_\ast \geq p_\ast \geq h_\ast$. In this step, we  prove   $q_\ast=H_\ast$ and $p_\ast=h_\ast$.

We use a contradiction argument. Let us assume for the moment  $p_\ast > h_\ast$ in order to eventually build a contradiction.
\\
Fix a time $s$ and take a node $i_0$  with $\mathbf{x}_{i_0}(s)=\min_{j}\mathbf{x}_j(s) \leq h_\ast< p_\ast$. Such a node always exists in view of the fact that $p_\ast > h_\ast=\lim_{t\to \infty} \min\{ \min_j \mathbf{x}_j(t), p_\ast\}$. The graph is strongly connected, therefore there must exist a node $i_1\neq i_0$ that is  influenced by node $i_0$ in the interaction graph, resulting in
\begin{align}
\label{eq:9}
\mathbf{x}_{i_1}(s+1)&=(1-\epsilon \sum_{j\in\mathrm{N}_{i_1}} a_{i_1j})\mathbf{x}_{i_1}(s)+ \epsilon \sum_{j\in\mathrm{N}_{i_1}} a_{i_1j} \psi_j (\mathbf{x}_j(s)) \nonumber\\
&\quad=(1-\epsilon \sum_{j\in\mathrm{N}_{i_1}} a_{i_1j})\mathbf{x}_{i_1}(s)+ \epsilon \sum_{j\neq i_0\in\mathrm{N}_{i_1}}a_{i_1j} \psi_j (\mathbf{x}_j(s))
 +\epsilon a_{i_1i_0} \psi_{i_0} (\mathbf{x}_{i_0}(s))\nonumber\\
&\quad\leq (1-\epsilon \sum_{j\in\mathrm{N}_{i_1}} a_{i_1j})\mathbf{I}_{q_{\ast}}^+(\max_{j\in\mathrm{V}} \mathbf{x}_j(s))+  \epsilon \sum_{j\neq i_0\in\mathrm{N}_{i_1}}a_{i_1j} \mathbf{I}_{q_{\ast}}^+(\max_{j\in \mathrm{V}} \mathbf{x}_j(s)) + \epsilon a_{i_1i_0} p_\ast \nonumber\\
& \quad=  (1-\epsilon a_{i_1 i_0}) \mathbf{I}_{q_{\ast}}^+(\max_{j\in \mathrm{V} }\mathbf{x}_j(s)) + \epsilon a_{i_1i_0} p_\ast\nonumber\\
&\quad\leq (1-\theta)\mathbf{I}_{q_{\ast}}^+(\max_{j\in \mathrm{V} }\mathbf{x}_j(s)) +\theta p_\ast
\end{align}
for
\begin{equation*}
	\theta=\big\{\epsilon \min_{(i,j)\in\mathrm{E}}\{a_{ij}:i\neq j, a_{ij}\neq 0\},\, \min_i \{1-\epsilon \sum_{j\in\mathrm{N}_i} a_{ij}\}\big\},
\end{equation*}
where in the first inequality we have used
\begin{align*}
\mathbf{x}_{i_1}(s)&\leq \mathbf{I}_{q_{\ast}}^+(\max_{j\in\mathrm{V}} \mathbf{x}_j(s))\\
\psi_j (\mathbf{x}_j(s))&\leq \mathbf{I}_{q_{\ast}}^+(\max_{j\in\mathrm{V}} \mathbf{x}_j(s))\\
\psi_{i_0} (\mathbf{x}_{i_0}(t))&\leq \psi_{i_0} (h_\ast)\leq p_\ast;
\end{align*}
and the second inequality is due to the facts that $\mathbf{I}_{q_{\ast}}^+(\max_{j\in\mathrm{V}} \mathbf{x}_j(s)) \geq p_\ast$ and $\epsilon a_{i_1i_0} \geq \theta$.

On the other hand, for node $i_0$, we have
\begin{align}
\label{eq:9i0}
\mathbf{x}_{i_0}(s+1)
&=(1-\epsilon \sum_{j\in\mathrm{N}_{i_0}} a_{i_0j})\,\mathbf{x}_{i_0}(s)+ \epsilon \sum_{j\in\mathrm{N}_{i_0}} a_{i_0j} \psi_j (\mathbf{x}_j(s)) \nonumber\\
&\quad\le(1-\epsilon \sum_{j\in\mathrm{N}_{i_0}} a_{i_0j})\,p_\ast+ (\epsilon \sum_{j\in\mathrm{N}_{i_0}} a_{i_0j} )\,\mathbf{I}_{q_{\ast}}^+(\max_{j\in\mathrm{V}} \mathbf{x}_j(s))\notag\\
&\quad \le \theta \,p_\ast+(1-\theta)\, \mathbf{I}_{q_{\ast}}^+(\max_{j\in\mathrm{V}} \mathbf{x}_j(s))
\end{align}
Therefore, for $k=i_0,i_1$, we have
\begin{equation*}
\mathbf{x}_{k}(s+1)
\le \theta \,p_\ast+(1-\theta)\, \mathbf{I}_{q_{\ast}}^+(\max_{j\in\mathrm{V}} \mathbf{x}_j(s))
\end{equation*}
Continuing to investigate time instant $s+2$, we have
\begin{align}
\label{eq:9_s+2}
&\mathbf{x}_{k}(s+2)
=(1-\epsilon \sum_{j\in\mathrm{N}_k} a_{kj})\,\mathbf{x}_k(s+1) +   \epsilon \sum_{j\in\mathrm{N}_k} a_{kj} \psi_j (\mathbf{x}_j(s+1)) \notag\\
&\quad\le(1-\epsilon \sum_{j\in\mathrm{N}_k} a_{kj})\,
\left[ \theta \,p_\ast+(1-\theta)\, \mathbf{I}_{q_{\ast}}^+(\max_{j\in\mathrm{V}} \mathbf{x}_j(s)) \right]   + (\epsilon \sum_{j\in\mathrm{N}_k} a_{kj}) \,\mathbf{I}_{q_{\ast}}^+(\max_{j\in\mathrm{V}} \mathbf{x}_j(s)) \notag\\
&\quad=\theta \,p_\ast+ \theta (\epsilon \sum_{j\in\mathrm{N}_k} a_{kj})\,
	(\mathbf{I}_{q_{\ast}}^+(\max_{j\in\mathrm{V}} \mathbf{x}_j(s))- p_\ast)  + (1-\theta)\, \mathbf{I}_{q_{\ast}}^+(\max_{j\in\mathrm{V}} \mathbf{x}_j(s)) \notag\\
&\quad\le\theta \,p_\ast+ \theta(1-\theta)\,
	(\mathbf{I}_{q_{\ast}}^+(\max_{j\in\mathrm{V}} \mathbf{x}_j(s))- p_\ast) \nonumber\\
&\ \ \ \ \ +  (1-\theta)\, \mathbf{I}_{q_{\ast}}^+(\max_{j\in\mathrm{V}} \mathbf{x}_j(s)) \notag\\
&\quad=\theta^2 \,p_\ast+(1-\theta^2)\, \mathbf{I}_{q_{\ast}}^+(\max_{j\in\mathrm{V}} \mathbf{x}_j(s))
,\quad k=i_0,i_1
\end{align}
This recursion gives us
\begin{align}
\label{eq:9recursion}
\mathbf{x}_{k}(s+\tau)
=\theta^\tau p_\ast+(1-\theta^\tau)\mathbf{I}_{q_{\ast}}^+(\max_{j\in\mathrm{V}} \mathbf{x}_j(s))
\end{align}
for $k=i_0,i_1,\;\tau=1,\dots,n-1$.
Note that $i_2$ is influenced by either $i_0$ or $i_1$, and without loss of generality we assume it is $i_1$ that is affecting $i_2$.
Then
\begin{align}
\label{eq:10}
\mathbf{x}_{i_2}(s+2)
	&=(1-\epsilon \sum_{j\in\mathrm{N}_{i_2}} a_{i_2j})\mathbf{x}_{i_2}(s+1)+\epsilon \sum_{j\neq i_1\in\mathrm{N}_{i_2}}a_{i_2j} \psi_j (\mathbf{x}_j(s+1)) +\epsilon a_{i_2i_1} \psi_{i_1} (\mathbf{x}_{i_1}(s+1))\notag\\
&\quad\leq (1-\epsilon a_{i_2i_1})\mathbf{I}_{q_{\ast}}^+(\max_{j\in\mathrm{V}} \mathbf{x}_j(s))+ \epsilon a_{i_2i_1} \,\left[  (1-\theta) \mathbf{I}_{q_{\ast}}^+(\max_{j\in\mathrm{V}} \mathbf{x}_j(s))+ \theta p_\ast\right] \notag\\
&\quad\leq (1-\theta)\,\mathbf{I}_{q_{\ast}}^+(\max_{j\in\mathrm{V}} \mathbf{x}_j(s))+ \theta \,\left[(1-\theta) \mathbf{I}_{q_{\ast}}^+(\max_{j\in\mathrm{V}} \mathbf{x}_j(s))+ \theta p_\ast\right] \notag\\
&\quad= (1-\theta^2)\,\mathbf{I}_{q_{\ast}}^+(\max_{j\in\mathrm{V}} \mathbf{x}_j(s)) +\theta^2 p_\ast
\end{align}
A similar recursion leads to
\begin{align}
\label{eq:10recursion}
\mathbf{x}_{i_2}(s+\tau)\le (1-\theta^\tau)\,\mathbf{I}_{q_{\ast}}^+(\max_{j\in\mathrm{V}} \mathbf{x}_j(s)) +\theta^\tau p_\ast
\end{align}
for $\tau=2,\dots,n-1$.
The strong connectivity of the graph allows us to continue the process until all nodes are visited, leading to
\begin{align}
\label{eq:11}
\mathbf{x}_{k}(s+n-1)\le (1-\theta^{n-1})\,\mathbf{I}_{q_{\ast}}^+(\max_{j\in\mathrm{V}} \mathbf{x}_j(s)) +\theta^{n-1} p_\ast
\end{align}
for $k=i_0,\dots,i_{n-1}$,
and thus
\begin{align}
\label{eq:12}
\max_{j\in\mathrm{V}} \mathbf{x}_{j}(s+n-1)\le  (1-\theta^{n-1})\,\mathbf{I}_{q_{\ast}}^+(\max_{j\in\mathrm{V}} \mathbf{x}_j(s)) +\theta^{n-1} p_\ast
\end{align}
with $k=i_0,\dots,i_{n-1}$.

At this point we investigate two cases, respectively.
\begin{itemize}
	\item[(a)] Let $p_\ast < H_\ast$. In this case,  for sufficiently large $s$, $\mathbf{I}_{q_{\ast}}^+(\max_{j\in\mathrm{V}} \mathbf{x}_j(s))$ will be so close to $H_\ast$ that
	\begin{equation*}
	(1-\theta^{n-1})\mathbf{I}_{q_{\ast}}^+(\max_{j\in\mathrm{V}} \mathbf{x}_j(s)) +\theta^{n-1} p_\ast<H_\ast.
	\end{equation*}
	Therefore, \eqref{eq:12} implies that
	\begin{align}
	\max_{j\in\mathrm{V}} \mathbf{x}_{j}(s+n-1)< H_\ast
	\end{align}
	for all $s$ that are sufficiently large.
	From the definition of $H(\mathbf{x}(t))$ and $H_\ast$, we can only conclude  $H_\ast =q_\ast$. As a result,  $\max_{j\in\mathrm{V}} \mathbf{x}_{j}(s+n-1)< q_\ast$ for sufficiently large $s$, which implies that there exists $T>0$ such that
	\begin{equation*}
		\psi_j(\mathbf{x}_j(t))\leq \mathbf{I}_{p_\ast}^+(\max_{j\in\mathrm{V}} x_j(t))
	\end{equation*}
	for all $j$ and all $t\geq T$.
	
	This means, the term $\mathbf{I}_{q_{\ast}}^+(\max_{j\in\mathrm{V}} \mathbf{x}_j(s))$ in Eqs \eqref{eq:9}, \eqref{eq:10}, \eqref{eq:11} can be replaced by $\mathbf{I}_{p_{\ast}}^+(\max_{j\in\mathrm{V}} \mathbf{x}_j(s))$ for $s>T$. In this case \eqref{eq:12} becomes
	\begin{align}
	\max_{j\in\mathrm{V}} \mathbf{x}_{j}(s+n-1)&\leq (1-\theta^{n-1})\mathbf{I}_{p_{\ast}}^+(\max_{j\in\mathrm{V}} \mathbf{x}_j(s)) \nonumber\\
&\ \ \ +\theta^{n-1} p_\ast
	\end{align}
	for all  $s\geq T$. Letting $s$ tend to infinity from both sides of the inequality we know   $$\limsup_{t\to \infty} \max_{j\in\mathrm{V}} \mathbf{x}_{j}(t)\leq p_\ast.
	$$
	\item[(b)] Suppose $p_\ast = H_\ast$. Then of course $\limsup_{t\to \infty} \max_{j\in\mathrm{V}} \mathbf{x}_{j}(t)\leq p_\ast=H_\ast$.
\end{itemize}

Therefore, there must hold true that $\limsup_{t\to \infty} \max_{j\in\mathrm{V}} \mathbf{x}_{j}(t)\leq p_\ast$. On the other hand, $p_\ast > h_\ast$ implies that there also holds true $\lim_{t\to \infty} \min_j \mathbf{x}_{j}(t)=h_\ast$. An immediate conclusion we can draw from the structure of the algorithm is that it can only be the case $\limsup_{t\to \infty} \max_{j\in\mathrm{V}} \mathbf{x}_{j}(t)= p_\ast$ because otherwise, there is a node $i_\ast$ with $\psi_{i_\ast}(\mathbf{x}_{i_\ast}(t))=p_\ast$ for all $t$ that are large enough. However, even $\limsup_{t\to \infty} \max_{j\in\mathrm{V}} \mathbf{x}_{j}(t)= p_\ast$ ensures that there must always be nodes whose  states  are arbitrarily close to $p_\ast$ for an infinite amount of times, a similar contradiction argument would clarify that in that case $\lim_{t\to \infty} \min_j \mathbf{x}_{j}(t)=p_\ast$ holds as well. This contradicts our standing assumption $p_\ast > h_\ast$.

We have now proved $p_\ast = h_\ast$. A symmetric  argument leads to $q_\ast = H_\ast$ as well.

\medskip

\noindent {\bf Step 3}. We rewrite  the update of node $i$ as
\begin{align}
\mathbf{x}_i(t+1)=(1-\epsilon \sum_{j \in\mathrm{N}_i} a_{ij})\mathbf{x}_i(t)+ \epsilon\sum_{j\in\mathrm{N}_i}  a_{ij}  \mathbf{x}_j(t) +w_i(t)
\end{align}
with
$
w_i(t):=\epsilon\sum_{j\in\mathrm{N}_i} a_{ij}\big(\psi_j(\mathbf{x}_j(t))-\mathbf{x}_j(t)\big).
$
Then we can reach
\begin{align}
\limsup_{t\rightarrow +\infty} \max_{i,j\in\mathrm{V}}\big |\mathbf{x}_i(t)-\mathbf{x}_j(t)\big|=0.
\end{align}
by the  robust consensus results for discrete-time dynamics  \cite{wang-liu-2009}. The final piece of proof for node state convergence follows from the same argument as the proof for continuous-time dynamics, and then we finally have  $\lim_{t \to \infty} \mathbf{x}_i(t)=c^\ast$ for all $i$ with $c^\ast \in[p_\ast,q_\ast]$. This completes the proof. \hfill$\square$

\begin{figure*}[ht]
\centering
\subfigure[]{
\includegraphics[trim=0cm 0cm 0cm 0cm,clip=true,width=7cm]{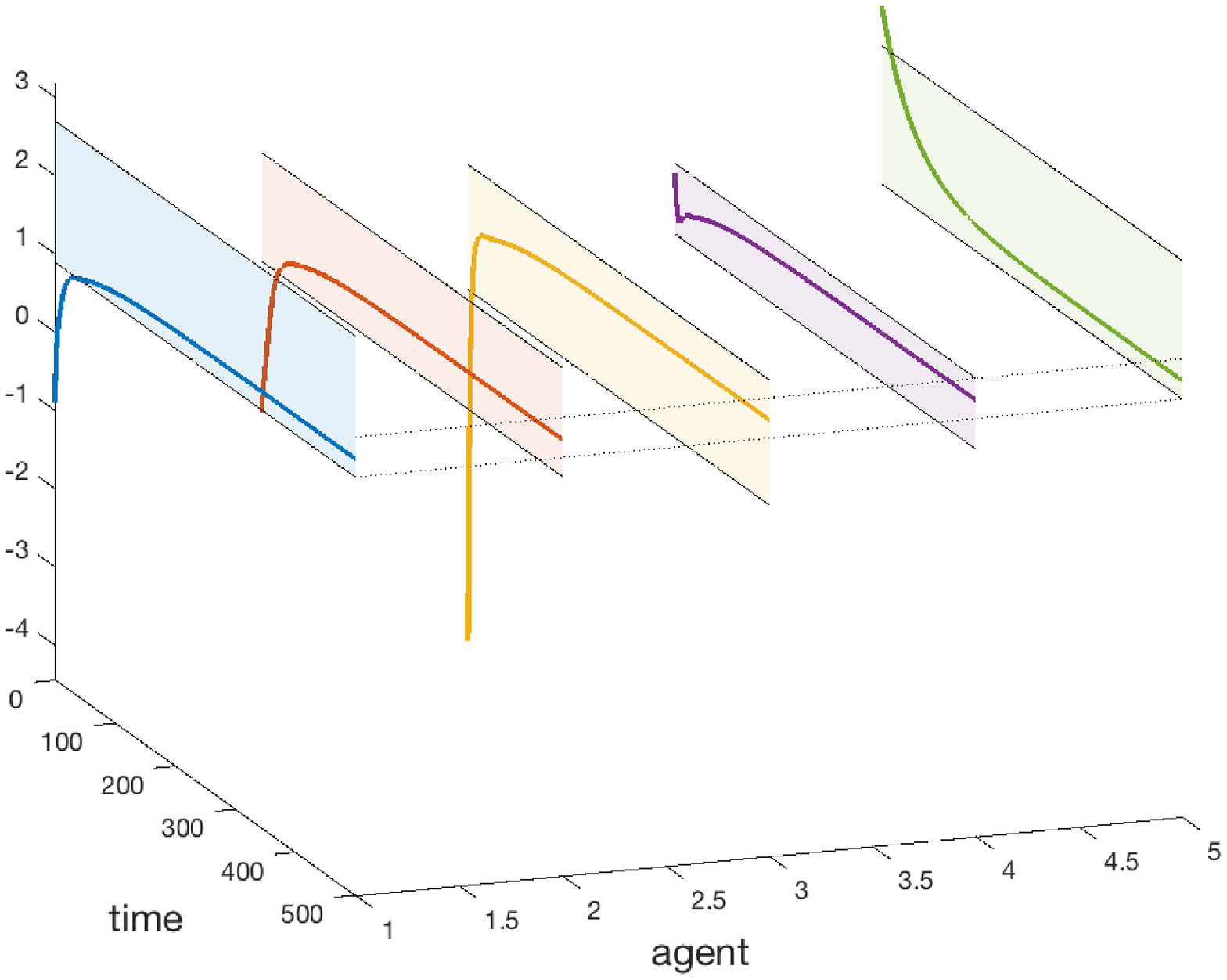}}
\subfigure[]{%\subfigure[]{
\includegraphics[trim=0cm 0cm 0cm 0cm,clip=true,width=7cm]{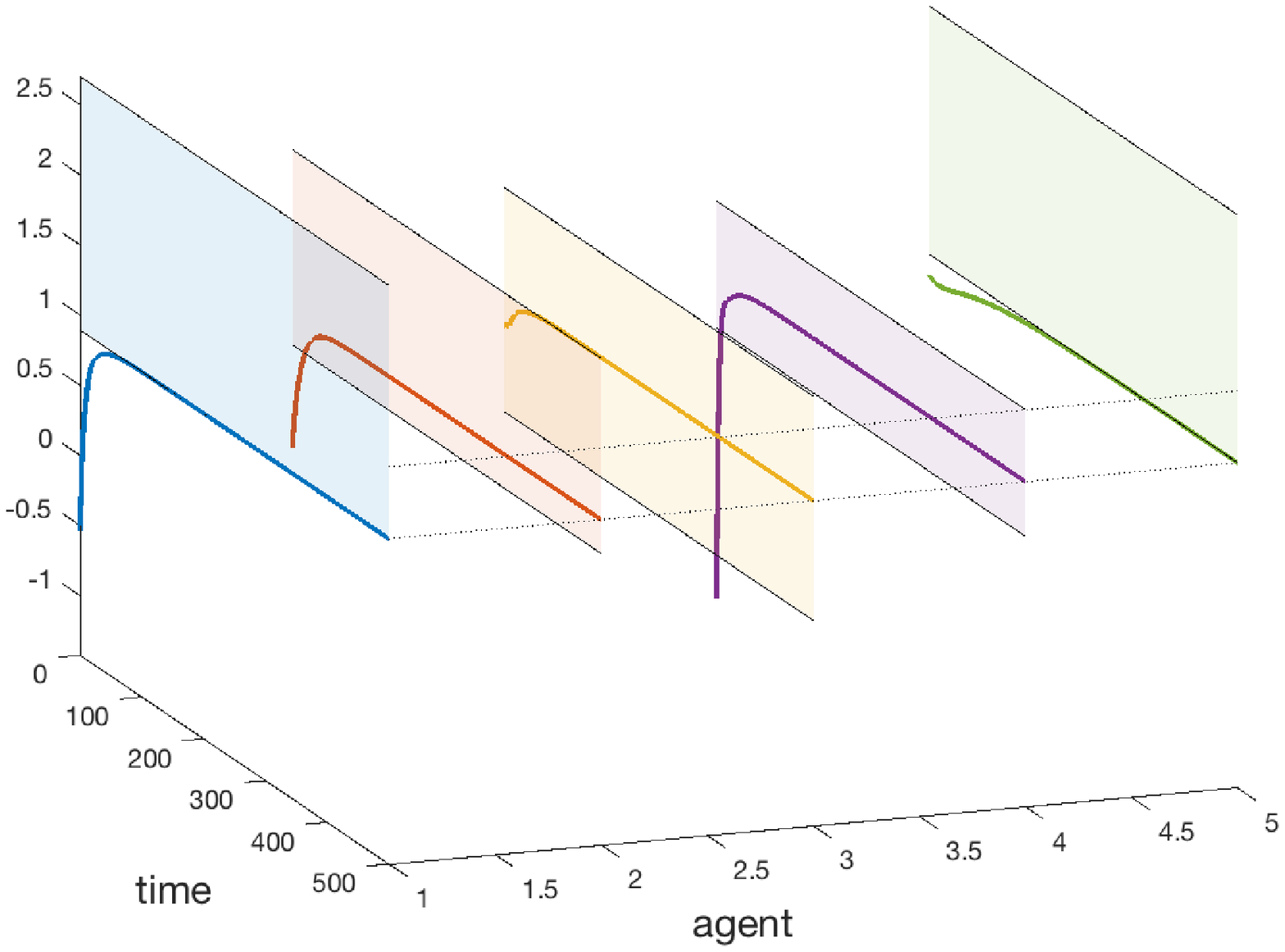}} \\
\subfigure[]{%\subfigure[]{
\includegraphics[trim=0cm 0cm 0cm 0cm,clip=true,width=7cm]{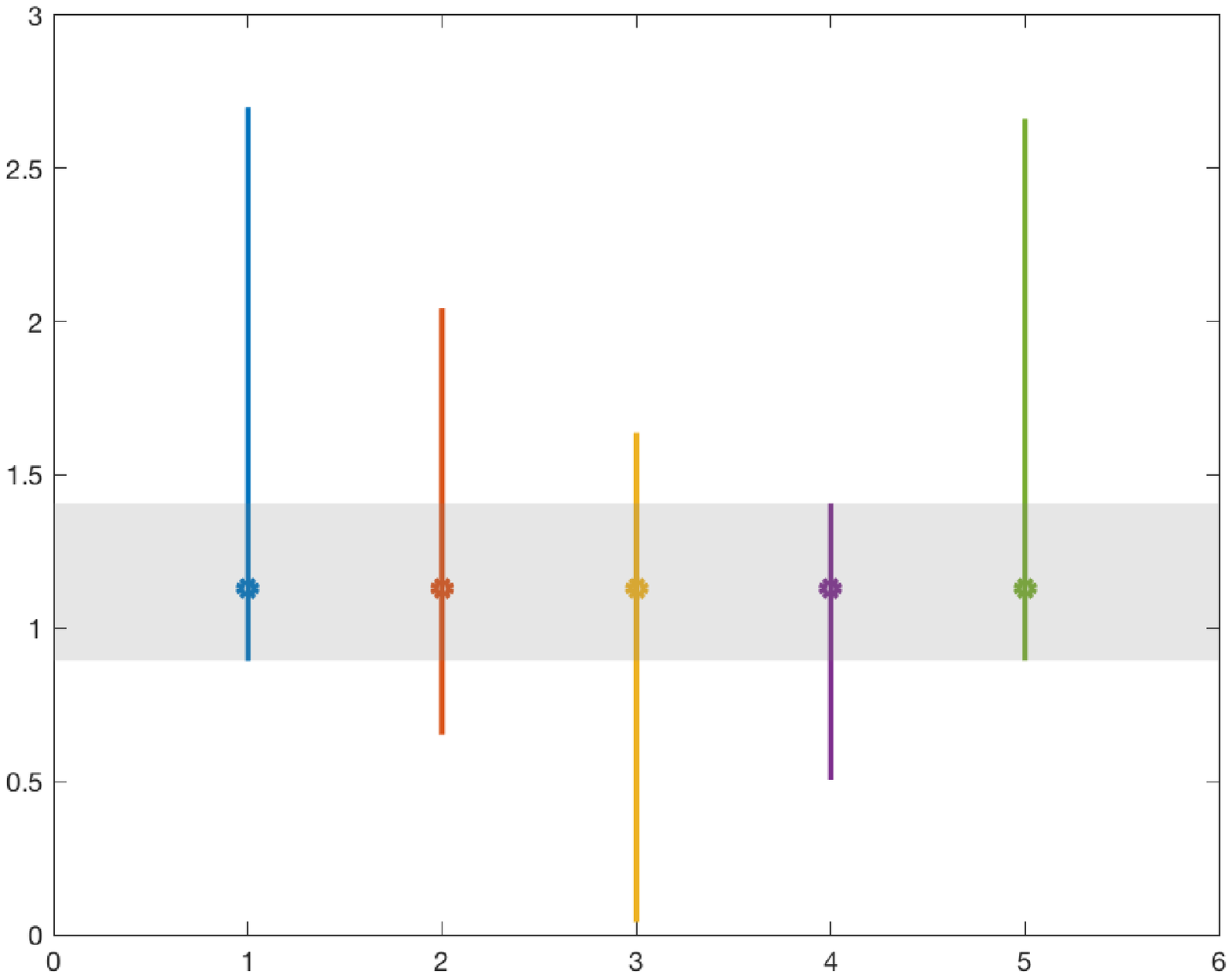}}
\subfigure[]{%\subfigure[]{
\includegraphics[trim=0cm 0cm 0cm 0cm, clip=true,  width=7cm]{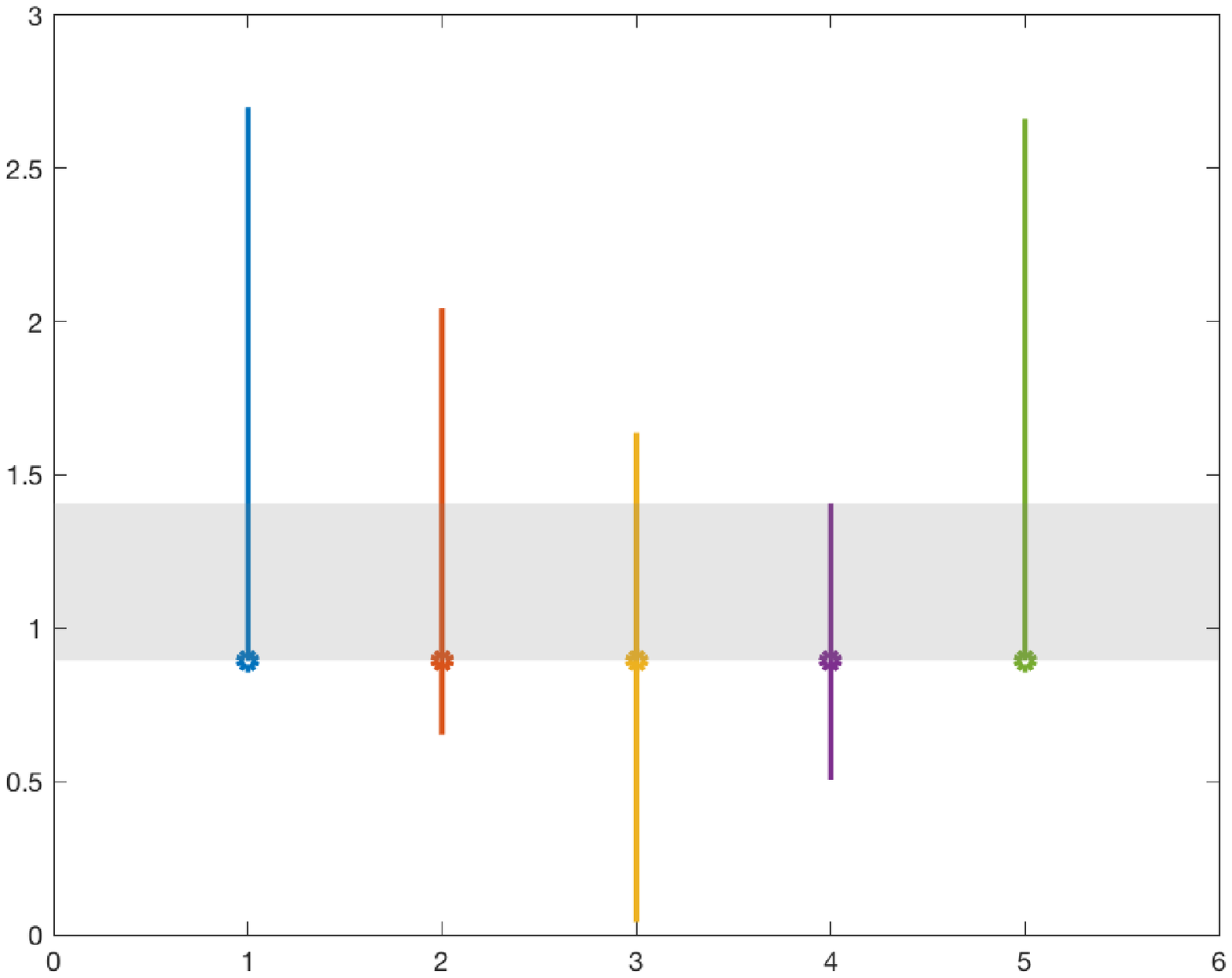}}
\caption{\small Two simulations for Example~\ref{example:counter1}, from different initial conditions. Top row: the trajectories $ \mathbf{x} (t)$ of the agents are shown in solid color lines. For each agent the shaded region represents the intervals $ [p_m, \, q_m ] $, while the transversal dotted lines are $ p_\ast$ and $ q_\ast$.
Bottom row: Intervals $ [p_m, \, q_m ] $ for each of the 5 agents, and consensus value $ c^\ast $ (circle) are shown in color, while the grey shaded region corresponds to $ [p_\ast, \,  q_\ast]$. In the left column we have $ c^\ast \in ( p_\ast, \, q_\ast ) $, while in the right column $ c^\ast = p_\ast$.}
\label{fig:simulation}
\end{figure*}

\section{Numerical Examples}
\label{sec:examples}

In this section we first consider a case in which the intervals $\mathcal{I}_m$ have nonempty intersection, and then an empty intersection case. Our third example is a cycle graph also with empty interval intersection for which the equilibrium point can be computed explicitly.

\begin{figure*}[ht]
\centering
\subfigure[]{
\includegraphics[trim=0cm 0cm 0cm 0cm,clip=true,width=5cm]{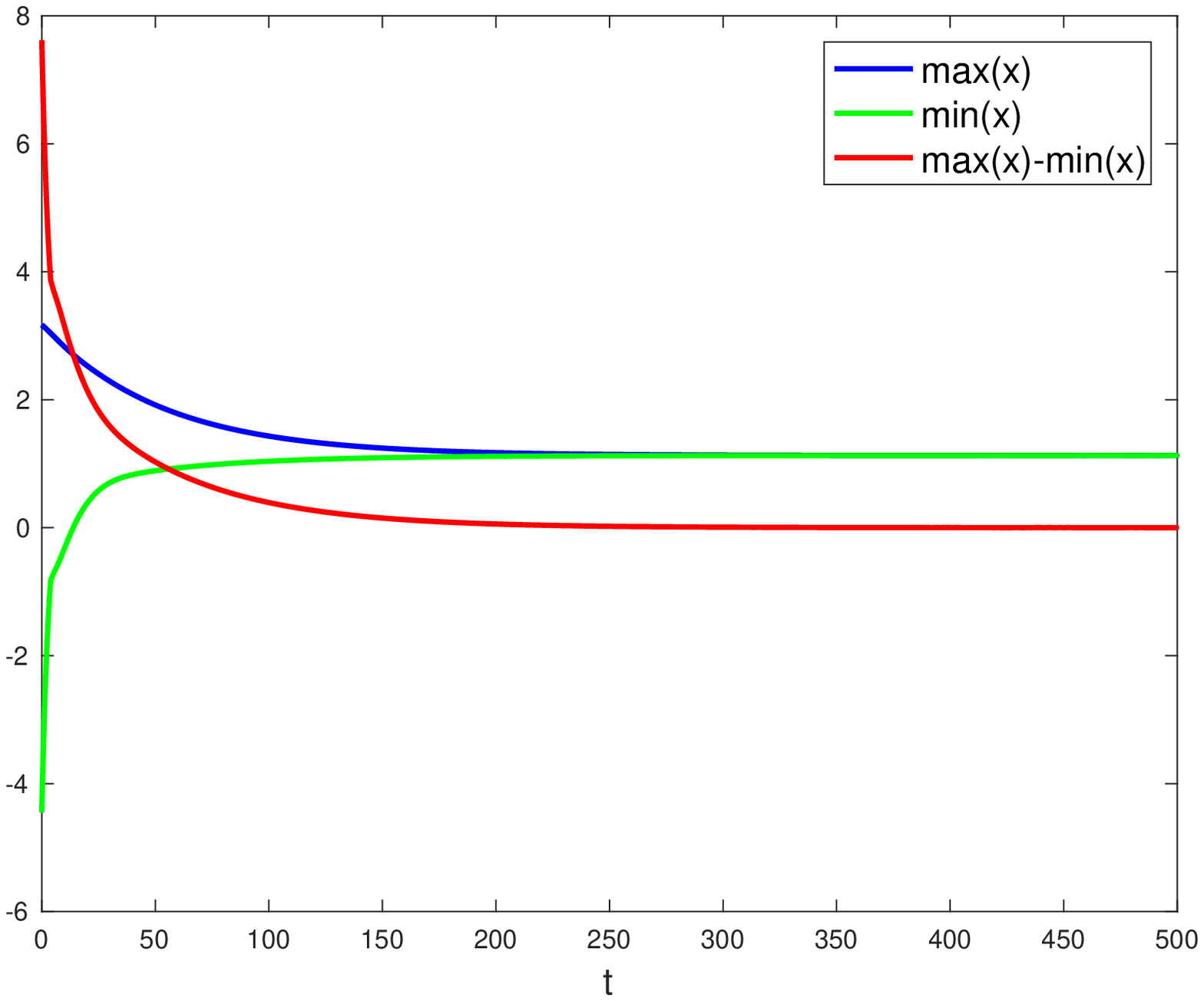}} $ \qquad \qquad $
\subfigure[]{%\subfigure[]{
\includegraphics[trim=0cm 0cm 0cm 0cm,clip=true,width=5cm]{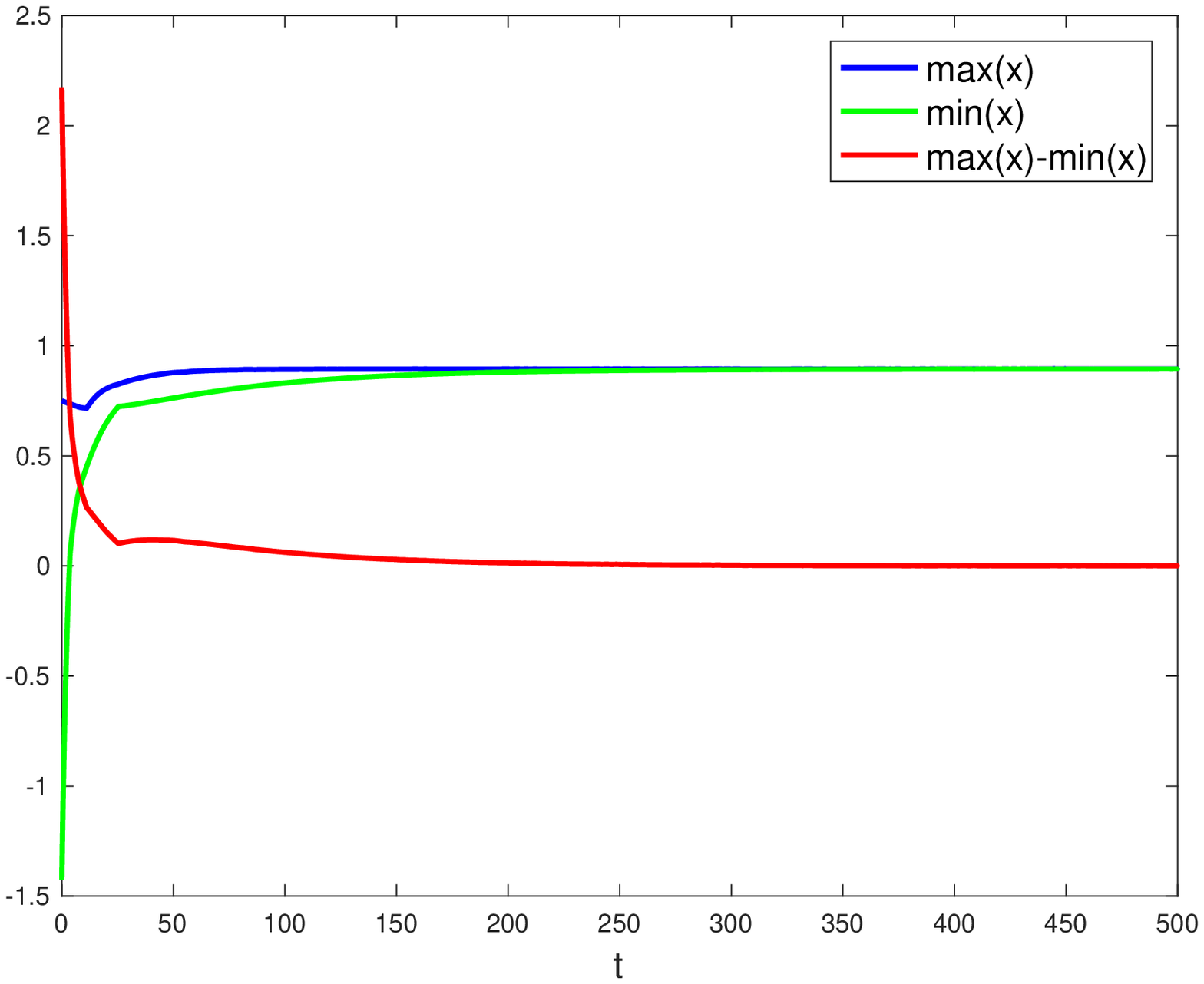}}
\caption{\small Values of $ \max_i \{\mathbf{x}_i(t)\}$ (blue), $ \min_i \{\mathbf{x}_i(t)\}$ (green), and $ \max_i \{\mathbf{x}_i(t)\} - \min_i \{\mathbf{x}_i(t)\}$ (red) for the simulations in Fig.~\ref{fig:simulation}. It can be seen that neither $ \max_i \{\mathbf{x}_i(t)\}$ nor $ \max_i \{\mathbf{x}_i(t)\} - \min_i \{\mathbf{x}_i(t)\}$ are monotone in the second case.}
\label{fig:simulation2}
\end{figure*}

\begin{example}
\label{example:counter1}
In Fig.~\ref{fig:simulation} an example of interval consensus with $ n=5$ is shown in which $\bigcap_{m=1}^n \mathcal I_m = [p_\ast, \, q_\ast] \ne \emptyset$. In the left column the consensus value $ c^\ast$ is strictly inside the interval $ [p_\ast, \, q_\ast]$.
In the right column instead $ c^\ast $ is on the boundary of $  [p_\ast, \, q_\ast] $ ($ c^\ast = p_\ast $) and it is clearly driven there by the saturation on $ \psi(\mathbf{x})$. Notice that, unlike for a standard consensus problem, in the process of converging neither $ \max_i \{\mathbf{x}_i(t)\}$ nor $ \max_i \{\mathbf{x}_i(t)\} - \min_i \{\mathbf{x}_i(t)\}$ are monotonically decreasing, see Fig.~\ref{fig:simulation2}. Notice further that $\mathbf{x}^0$ need not belong to $\prod_{i=1}^n [p_i,\,q_i]=[p_1,\,q_1]\times \dots \times [p_n,\,q_n]$, i.e., convergence is for any $\mathbf{x}^0\in \R^n$.
As the left column of Fig.~\ref{fig:simulation} shows, $ \mathbf{x}^0 \notin \prod_{i=1}^n [p_i,\,q_i]$ does not necessarily lead to $ c^\ast $ on the boundary of $  [p_\ast, \, q_\ast]$.
\end{example}

%---------------------------------------------------------------------------------

\begin{example}
\label{ex:example_nointersection}
In the $n=5 $ example of Fig.~\ref{fig:simulation_nointersection}, the intervals $\mathcal{I}_m$ have empty intersection, i.e., $\bigcap_{m=1}^n \mathcal I_m= \emptyset$.  Numerically the system \eqref{eq:system1} admits a unique equilibrium point which is not a consensus value, but which appears to be asymptotically stable in the entire $ \mathbb{R}^5$.

\begin{figure*}[ht]
\centering
\subfigure[]{
\includegraphics[trim=0cm 0cm 0cm 0cm,clip=true,width=6.5cm]{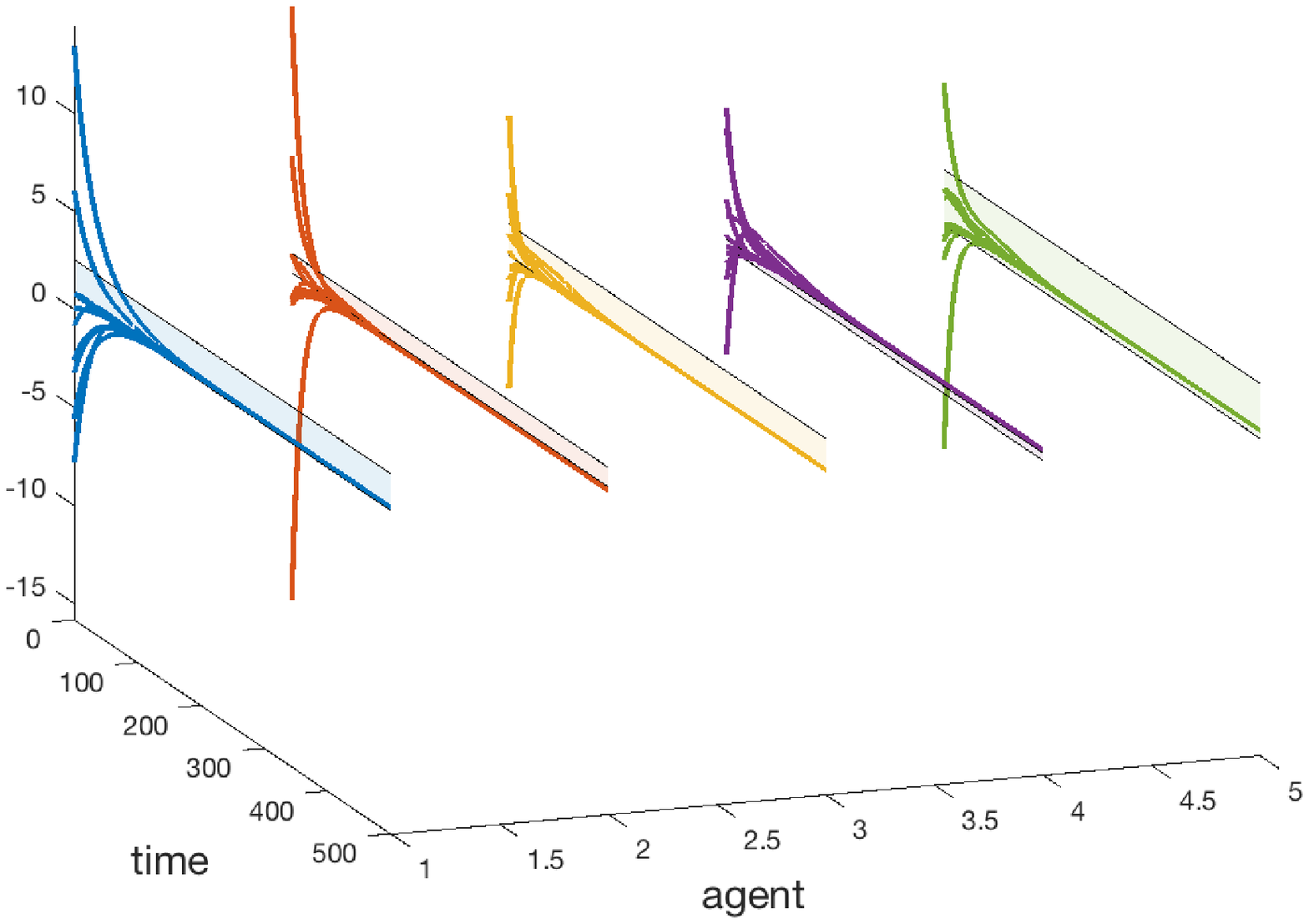}} $ \qquad \qquad $
\subfigure[]{%\subfigure[]{
\includegraphics[trim=0cm 0cm 0cm 0cm,clip=true,width=5.5cm]{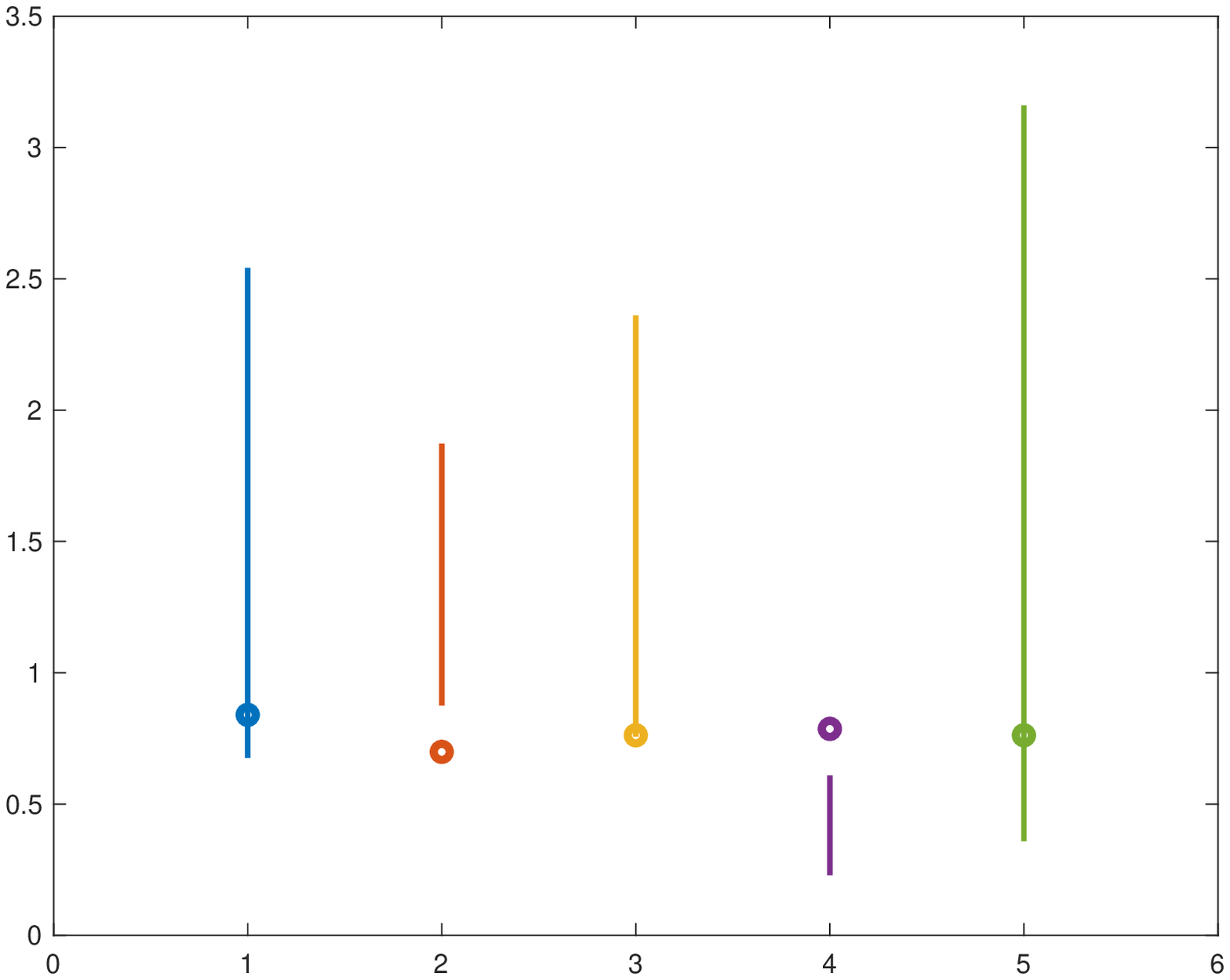}}
\caption{\small Example~\ref{ex:example_nointersection}. (a): Trajectories $ \mathbf{x}(t)$ of the system \eqref{eq:system1} from 10 random initial conditions. Shadowed region: $p_m$ and $q_m$ for each agent.
(b): Intervals $[p_m,q_m]$ for each agent and equilibrium $\mathbf{e}$ (circle).}
\label{fig:simulation_nointersection}
\end{figure*}

\end{example}

\begin{example}
\label{ex:example2}
In this example, we still consider empty intersection between the sets, i.e., $\bigcap_{m=1}^n \mathcal I_m=\emptyset$ or, equivalently, $q_\ast < p_\ast$, and in addition we assume that $p_1\le p_2\le\dots\le p_n$ and $q_1\le\dots\le q_n$.
If the graph is a (strongly connected) cycle graph (which has $|N_m|=1$ for all $m$, see Fig.~\ref{fig:example2} panel (a)), then it follows from Theorem~\ref{thm:cycle-convergence} that \eqref{eq:system1} admits a unique equilibrium point $ \mathbf{e}$ which is in $[q_\ast, p_\ast]$.
However, this special case is interesting because it is possible to compute $ \mathbf{e}$ in an explicit way, directly from the $ p_m $ and $ q_m$.
The adjacency matrix $A=[a_{ij}] $ has the following cyclic structure:
	\begin{equation*}
	a_{ij} =
	\begin{cases}
		a_{i,i+1}\ne 0,& j=i+1 \\
		0,& j\ne i+1
	\end{cases} \qquad
	i=1,\dots,n-1
	\end{equation*}
and
\[
a_{nj} =
	\begin{cases}
		a_{n1}\ne 0, & j=1 \\
		0, & j\ne 1
	\end{cases}.
\]
In this case \eqref{eq:system1} becomes
	\begin{align}
%		\frac{d}{dt}\mathbf{x}_1(t)&= a_{12}\Big(\psi_2\big(\mathbf{x}_2(t) \big)-\mathbf{x}_1(t) \Big)\\
%		\vdots&\\
		\frac{d}{dt}\mathbf{x}_i(t)&= a_{i,i+1}\Big(\psi_{i+1}\big(\mathbf{x}_{i+1}(t) \big)-\mathbf{x}_{i}(t) \Big)\quad i=1,\dots, n-1 \notag\\		
		\frac{d}{dt}\mathbf{x}_n(t)&= a_{n1}\Big(\psi_1\big(\mathbf{x}_1(t) \big)-\mathbf{x}_n(t) \Big)
	\label{eqn:systemEXCyclic}
	\end{align}
	and, from Theorem \ref{theorem:equilibrium}, it admits at least one equilibrium point, which is in $[\underline{p},\overline{q}]=[p_1,q_n]$.
	Let $\mathbf{e}$ be an equilibrium point of \eqref{eqn:systemEXCyclic}, that is
	\[
	\begin{cases}
	\mathbf{e}_{i}= \psi_{i+1}\big(\mathbf{e}_{i+1}\big), \quad i=1,\dots, n-1\\		
	\mathbf{e}_n= \psi_1\big(\mathbf{e}_1 \big)
	\end{cases}
	\]
	From Theorem \ref{theorem:equilibrium}, we know that $\mathbf{e}_1\geq p_1$, which implies that $\psi_1\big(\mathbf{e}_1 \big)=\mathbf{e}_1$ if $\mathbf{e}_1\le q_1$ or $\psi_1\big(\mathbf{e}_1 \big)=q_1$ if $\mathbf{e}_1>q_1$. Then
	\begin{equation*}
	\mathbf{e}_n=
	\begin{cases}
		\mathbf{e}_1, &  \text{if } \mathbf{e}_1 \le q_1\\
		q_1, &  \text{if } \mathbf{e}_1>q_1		
	\end{cases}
	\end{equation*}
	and
	\begin{equation*}
	\mathbf{e}_{n-1}=
	\begin{cases}
		\psi_n\big(\mathbf{e}_1\big) , & \text{if } \mathbf{e}_1 \le q_1\\
		\psi_n\big(q_1\big) , &  \text{if } \mathbf{e}_1>q_1
	\end{cases}
	\quad =
	\begin{cases}
		p_n , &  \text{if } \mathbf{e}_1 \le q_1\\
		p_n , &  \text{if } \mathbf{e}_1>q_1
	\end{cases}
	\quad =p_n
	\end{equation*}
	because $q_1=q_\ast<p_\ast =p_n$.
	Therefore
	\begin{equation*}
	\mathbf{e}_{n-2}= \psi_{n-1}\big(p_n\big)
	=
	\begin{cases}
		p_n , & \text{if } p_n\le q_{n-1}\\
		q_{n-1} , &  \text{if } p_n>q_{n-1}
	\end{cases}
	\end{equation*}
	because $p_n>p_{n-1}$, and
	\begin{align*}
	\mathbf{e}_{n-3}= \psi_{n-2}\big(\mathbf{e}_{n-2}\big)
	&=
	\begin{cases}
		\psi_{n-2}\big(p_n\big) , & \text{if } p_n\le q_{n-1}\\
		\psi_{n-2}\big(q_{n-1}\big) , & \text{if } p_n>q_{n-1}
	\end{cases}
	\nonumber\\
&=
	\begin{cases}
		q_{n-2} , & \text{if } p_n \in  (q_{n-2},q_{n-1}]\\
		p_n , & \text{if } p_n\le  q_{n-2}\\
		q_{n-2} , & \text{if } p_n>q_{n-1}
	\end{cases}
	\\&=
	\begin{cases}
		q_{n-2} , & \text{if } p_n > q_{n-2}\\
		p_n , & \text{if } p_n\le  q_{n-2}
	\end{cases}
	\end{align*}
	Iterating yields
	\begin{equation*}
	\mathbf{e}_{n-i}
	=
	\begin{cases}
		q_{n-i+1} , &  \text{if } p_n > q_{n-i+1}\\
		p_n , & \text{if } p_n\le  q_{n-i+1}
	\end{cases}
	\qquad i=1,\dots, n-1
	\end{equation*}
	and in particular
	\begin{equation*}
	\mathbf{e}_1
	=
	\begin{cases}
		q_2 , & \text{if } p_n > q_2\\
		p_n , &  \text{if } p_n\le  q_2
	\end{cases}.
	\end{equation*}
	Since $q_2>q_1$ and $p_n>q_1$, it follows that $\mathbf{e}_1>q_1$ and hence that $\mathbf{e}_n=q_1$. In conclusion, the system \eqref{eq:system1} admits a unique equilibrium point $\mathbf{e}$ such that
	\begin{align*}
		\mathbf{e}_n &= q_1
		\\
		\mathbf{e}_{n-1} &= p_n
		\\
		\mathbf{e}_{n-i}
		&=
		\begin{cases}
			q_{n-i+1} , &  \text{if } p_n > q_{n-i+1}\\
			p_n , & \text{if } p_n\le  q_{n-i+1}
		\end{cases}
		\qquad i=2,\dots, n-1
	\end{align*}
From Theorem~\ref{thm:local-stability}, the equilibrium must be asymptotically stable.
Moreover, it must be $\mathbf{e}\in[q_1,p_n] = [q_\ast,p_\ast]$.
Fig.~\ref{fig:example2} shows the result for a cycle graph of size $n=10$ nodes and edges weight drawn from a uniform distribution. The asymptotic stability character of the unique equilibrium point is confirmed, see panel (b).

\begin{figure*}[ht]
		\centering
		\subfigure[]{
			\includegraphics[trim=0cm 0cm 0cm 0cm, clip=true,  width=7cm]{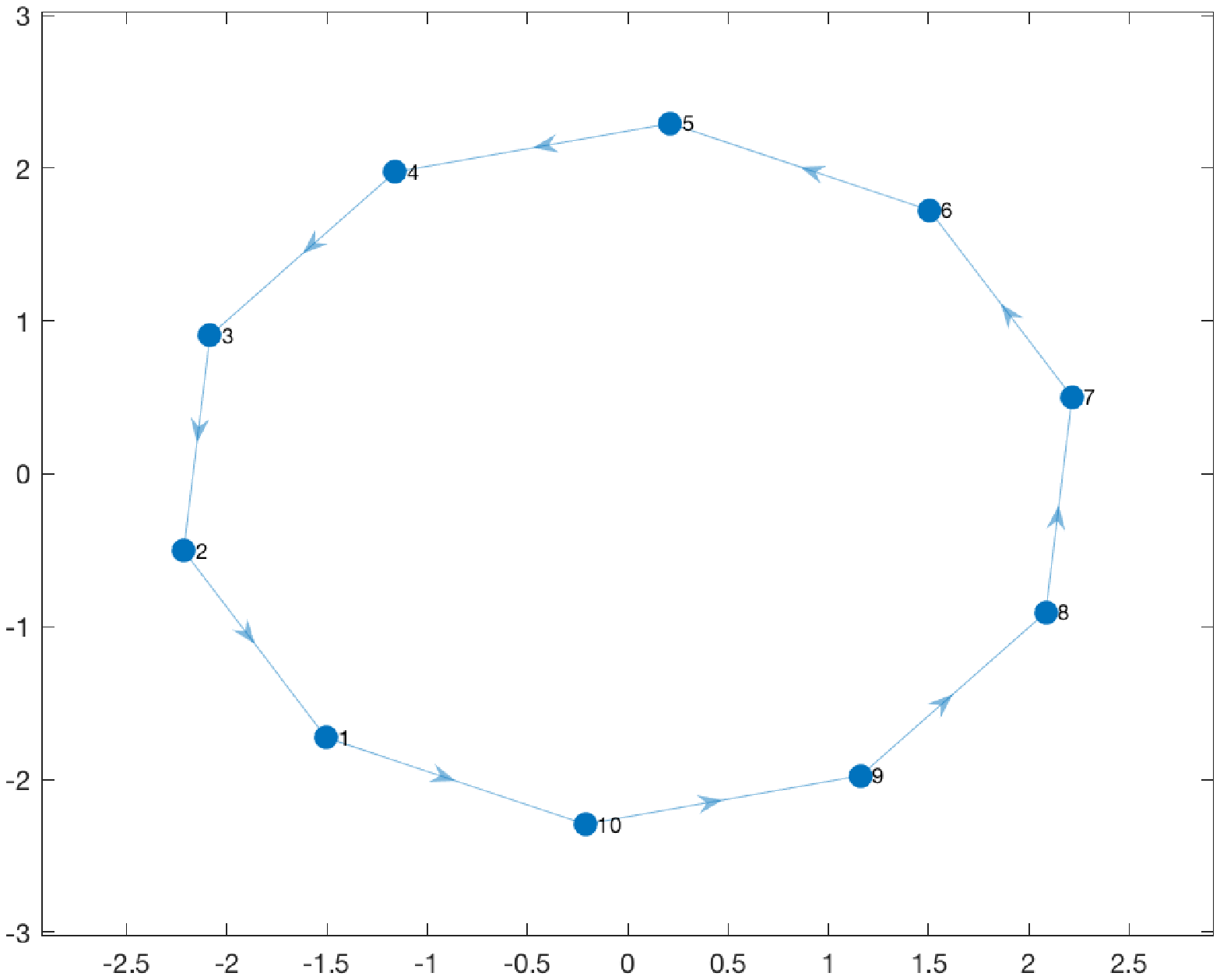}}
		\subfigure[]{
			\includegraphics[trim=0cm 0cm 0cm 0cm, clip=true,  width=7cm]{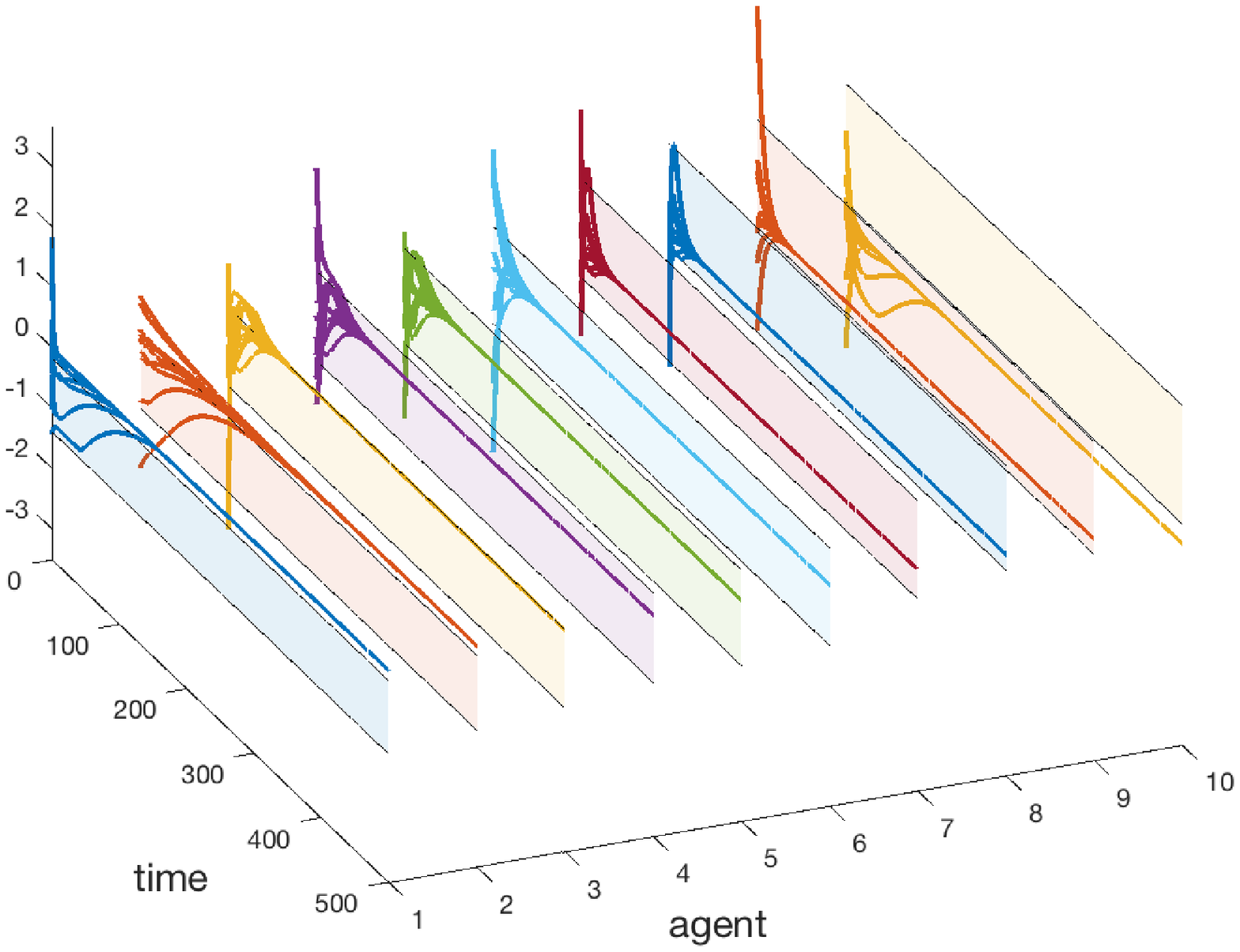}}
		\subfigure[]{
			\includegraphics[trim=0cm 0cm 0cm 0cm, clip=true,  width=6cm]{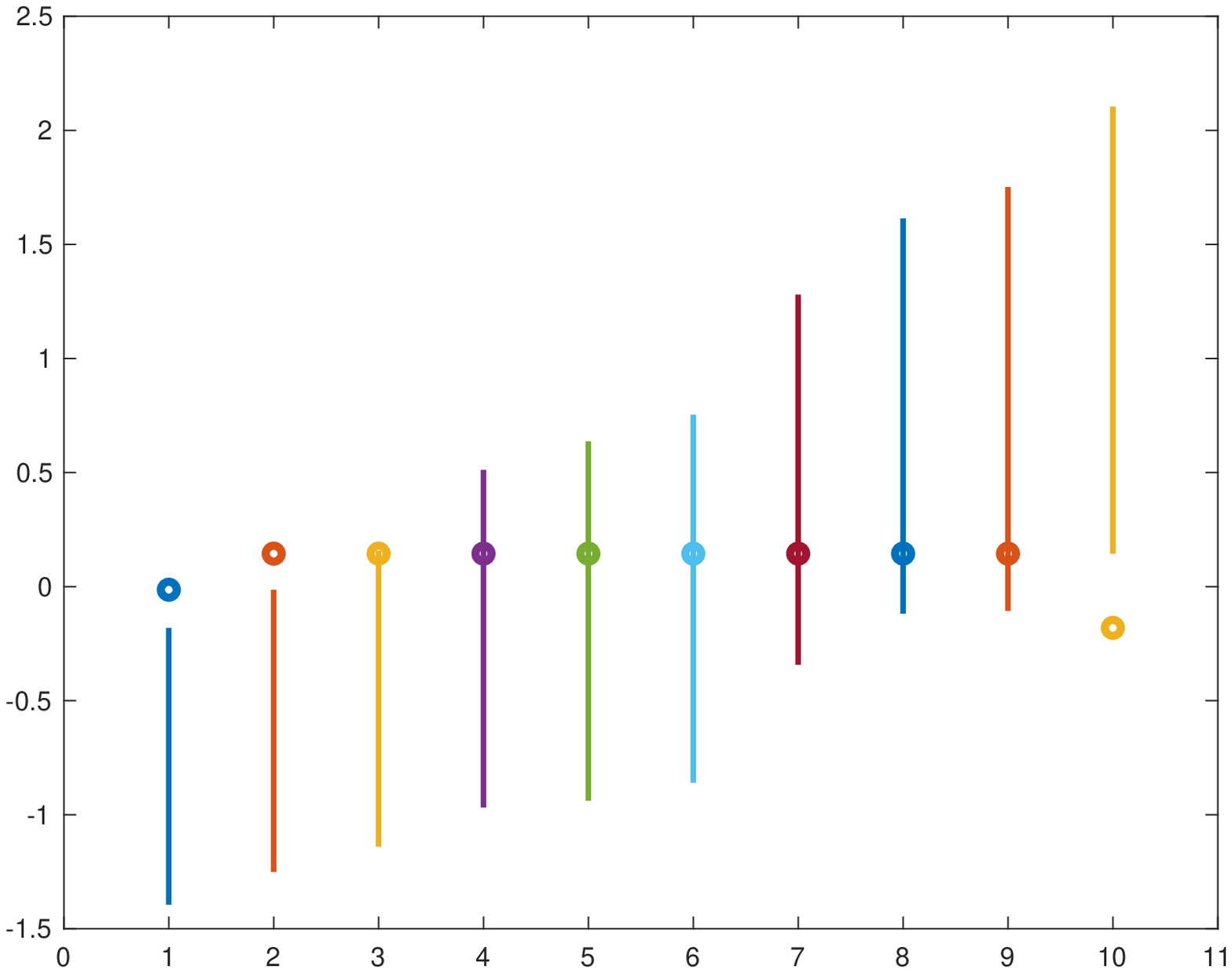}}
\caption{\small Example~\ref{ex:example2}.
(a): Cycle Graph.
(b): Trajectories $ \mathbf{x}(t)$ of the system \eqref{eq:system1} from 10 random initial conditions.
(c): Intervals $[p_m,q_m]$ for each agent and $\mathbf{e}$ (circle).}
		\label{fig:example2}
	\end{figure*}
\end{example}

\section{Conclusion}
If a group of agents seeking a consensus has non-dispensable requests on the range of values that such a consensus can achieve, then standard consensus algorithms cannot be used and something more sophisticated must be used.
The scheme proposed in this paper, interval consensus, allows to do this efficiently in both continuous and discrete-time with the only (unavoidable) prerequisite that the intersection of the agent intervals is nonempty.

To complete the understanding of our saturated dynamics \eqref{eq:system1} some work still need to be done for the cases with empty interval intersection. In particular, the conjecture which we could not fully prove, is that these cases always lead to a single (asymptotically stable) equilibrium point.

\end{document}